\documentclass[a4paper,fleqn,usenatbib]{mnras}

\usepackage[T1]{fontenc}
\usepackage{ae,aecompl}

\usepackage{graphicx}	
\usepackage{amsmath}	
\usepackage{amssymb}	

\newcommand{\Msun}{M_{\odot}}
\newcommand{\Rsun}{R_{\odot}}

\newcommand{\Mej}{M_{\rm ej}}

\newcommand{\Cofs}{$^{56}$Co}
\newcommand{\Nifs}{$^{56}$Ni}
\newcommand{\Ibar}{I\hspace{-.5em}{\scriptscriptstyle \stackrel{\mathbf{-}}{}}} 
\newcommand{\ergs}{erg~s$^{-1}$}

\title[Multi-messenger strategy based on CCSN simulation]{Multi-messenger signals of long-term core-collapse supernova simulations : synergetic observation strategies}

\author[K. Nakamura et al.]{
Ko Nakamura,$^{1,2}$\thanks{E-mail: nakamura.ko@heap.phys.waseda.ac.jp}
Shunsaku Horiuchi,$^{3}$\thanks{E-mail: horiuchi@vt.edu}
Masaomi Tanaka,$^{4}$
Kazuhiro Hayama,$^{5,4}$
\newauthor
Tomoya Takiwaki$^{4}$
and Kei Kotake$^{6,4}$
\\
$^{1}$Faculty of Science and Engineering, Waseda University, Ohkubo 3-4-1, Shinjuku, Tokyo 169-8555, Japan\\
$^{2}$Yukawa Institute for Theoretical Physics, Kyoto University, Kyoto 606-8502, Japan\\
$^{3}$Center for Neutrino Physics, Department of Physics, Virginia Tech, Blacksburg, Virginia 24061, USA\\
$^{4}$National Astronomical Observatory of Japan, Osawa 2-21-1, Mitaka, Tokyo 181-8588, Japan\\
$^{5}$KAGRA Observatory, Institute for Cosmic Ray Research, University of Tokyo,
238 Higashi Mozumi, Kamioka, Hida, Gifu 506-1205, Japan\\
$^{6}$Department of applied physics, Fukuoka University, Nanakuma Jonan 8-19-1, Fukuoka 814-0180, Japan
}

\date{Accepted 2016 June 15. Received 2016 June 14; in original form 2016 February 09}

\pubyear{2016}

\begin{document}
\label{firstpage}
\pagerange{\pageref{firstpage}--\pageref{lastpage}}
\maketitle

\begin{abstract}
The next Galactic supernova is expected to bring great opportunities for the direct detection of gravitational waves (GW), full flavor neutrinos, and multi-wavelength photons. To maximize the science return from such a rare event, it is essential to have established classes of possible situations and preparations for appropriate observations. To this end, we use a long-term numerical simulation of the core-collapse supernova (CCSN) of a 17 $M_\odot$ red supergiant progenitor to self-consistently model the multi-messenger signals expected in GW, neutrino, and electromagnetic messengers. This supernova model takes into account the formation and evolution of a protoneutron star, neutrino-matter interaction, and neutrino transport, all within a two-dimensional shock hydrodynamics simulation. With this, we separately discuss three situations: (i) a CCSN at the Galactic Center, (ii) an extremely nearby CCSN within hundreds of parsecs, and (iii) a CCSN in nearby galaxies within several Mpc. These distance regimes necessitate different strategies for synergistic observations. In a Galactic CCSN, neutrinos provide strategic timing and pointing information. We explore how these in turn deliver an improvement in the sensitivity of GW analyses and help to guarantee observations of early electromagnetic signals. To facilitate the detection of multi-messenger signals of CCSNe in extremely nearby and extragalactic distances, we compile a list of nearby red supergiant candidates and a list of nearby galaxies with their expected CCSN rates. By exploring the sequential multi-messenger signals of a nearby CCSN, we discuss preparations for maximizing successful studies of such an unprecedented stirring event. 
\end{abstract}




\section{Introduction}\label{sec:introduction}

Core-collapse supernovae (CCSNe) have been an active topic of research in theoretical and observational astrophysics for many decades. In recent years, hundreds of CCSNe have been discovered every year, mostly in distant galaxies \citep[e.g.,][]{Sako:2007ms,Leaman:2010kb}. However, these extragalactic CCSNe only provide a very indirect probe of the core-collapse process, and as a result, basic questions such as the mechanism by which CCSN explode still remains a mystery \citep[for reviews, see, e.g.,][]{Kotake:2005zn,Janka:2012wk}. CCSNe in our Milky Way and neighboring galaxies, while rare, provide the opportunity to study CCSNe with a suite of new observables, including neutrinos \citep[e.g.,][]{Scholberg:2012id}, gravitational waves \citep[GW; e.g.,][]{Ott09,Kotake13}, 
and nuclear gamma rays \citep[e.g.,][]{1987ApJ...322..215G,Horiuchi:2010kq}. These are poised to more directly reveal the secrets laying deep inside the stellar photosphere.

\begin{figure*}
\begin{center}
\includegraphics[scale=0.65, bb=0 220 600 600]{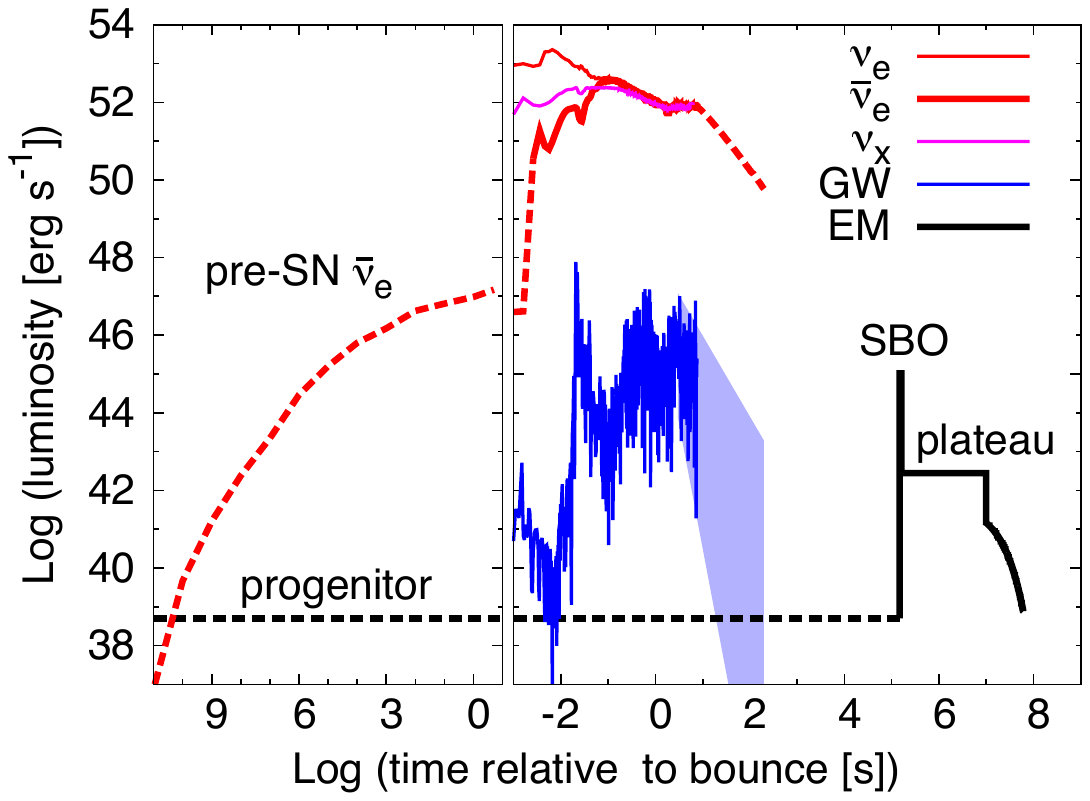}
\caption{
Time sequence for neutrino (red lines for $\nu_e$ and $\bar{\nu}_e$ and magenta line for $\nu_x$; $\nu_x$ represents heavy lepton neutrino $\nu_\mu$, $\nu_\tau$, $\bar{\nu}_\mu$, or $\bar{\nu}_\tau$), GW (blue line), and electromagnetic (EM, black line) signals based on our neutrino-driven core-collapse simulation of a non-rotating $17 \, \Msun$ progenitor. 
The solid lines are direct or indirect results of our CCSN simulation, whereas the dashed lines are from literatures or rough speculations.
The left (right) panel $x$-axis shows time before (after) core bounce. Emissions of pre-CCSN neutrinos as well as the core-collapse neutrino burst are shown as labeled. For the EM signal, the optical output of the progenitor, the SBO emission, the optical plateau, and the decay tail are shown as labeled. 
The GW luminosity is highly fluctuating during our simulation and the blue shaded area presents the region between the two straight lines fitting the high and low peaks during 3 -- 5 seconds postbounce. 
The hight of the curves does not reflect the energy output in each messenger; total energy emitted after bounce in the form of anti-electron neutrino, photons, and GW is $\sim 6 \times 10^{52}$ erg, $\sim 4 \times 10^{49}$ erg, and $\sim 7 \times 10^{46}$ erg, respectively. 
See the text for details. 
}
\label{fig:lumi}
\end{center}
\end{figure*}

Neutrinos and GWs in particular provide unique probes of the explosion mechanism in realtime. Multiple neutrino detectors capable of detecting CCSN neutrinos are currently in operation. The best suited is \textit{Super-Kamiokande} (Super-K) which is expected to collect a rich dataset of neutrino events from future Galactic CCSNe, while IceCube also has equal statistical detection potential even though IceCube cannot resolve individual neutrino events. Smaller detectors with sensitivity to CCSN neutrinos include, e.g., Baksan, Borexino, DayaBay, HALO, KamLAND, LVD, MiniBooNE, and NO$\nu$A \citep[for their detection potentials, see, e.g., recent review][]{Mirizzi:2015eza}. In the near-future, the \textit{Jiangmen Underground Neutrino Observatory} \citep[JUNO,][]{Li:2014qca} will augment Super-K and IceCube, and with future experiments such as \textit{Hyper-Kamiokande} \citep[Hyper-K,][]{Abe:2011ts} and \textit{Deep Underground Neutrino Experiment} \citep[DUNE,][]{Acciarri:2015uup}, neutrino event statistics and neutrino flavor information will be dramatically improved. GW detectors such as Advanced LIGO (aLIGO), Advanced Virgo (adVirgo), and KAGRA are expected to be able to detect CCSN GW out to a few kpc from the Earth, while future detectors such as the \textit{Einstein Telescope} (ET) can reach the entire Milky Way. 

In order to exploit these potentials, a multi-messenger observing strategy is necessary. In this context, the neutrino signal is particularly important. The neutrino emission in fact starts before the core collapse even begins. Neutrinos emitted during the final states of silicon burning can reach $\sim 5 \times 10^{50}$ erg for a massive star \citep{arnett89}, which can be detected by Hyper-K out to a few kpc away \citep{odrzywolek04}, thereby providing an early warning signal. During the first $\sim 10$ seconds after the core collapse, a copious $\sim 3\times 10^{53}$ erg of energy is emitted as neutrinos as was confirmed in SN~1987A \citep{hirata1987,Bionta:1987qt,Sato-and-Suzuki}. 

In addition to signaling unambiguously the occurrence of a nearby core collapse, the detected neutrinos will point to 
the location of the core collapse within an error circle of a few to ten degrees in the sky 
\citep{Beacom:1998fj,Tomas:2003xn,Bueno:2003ei}. This pointing information is particularly important for electromagnetic signals, which remain a crucial component of studies of CCSNe in the Milky Way and nearby galaxies. A few hours to days after the core collapse, the supernova shock breaks out of the progenitor surface, suddenly releasing the photons behind the shock in a flash bright in UV and X-rays, 
known as shock breakout (SBO) emission \citep{matzner99,blinnikov2000,tominaga2009,gezari2010,Kistler:2012as}. 
Although the SBO signal provides important information about the CCSN, such as the radius of the progenitor, detection is difficult because of its short duration. Knowing where to anticipate the signal will dramatically improve its detection prospects. In addition to the SBO, more traditional studies of CCSN properties (e.g, energy, composition, velocity) and its progenitor are important diagnostics of a CCSN, and a well-observed early light curve is important for accurate reconstruction of the CCSN evolution \citep[e.g.,][]{tominaga11}. 

\begin{table*}
\begin{center}
\caption{Detectable signals, detectors, and their horizons}\label{tbl:summary}
\begin{tabular}{clcccccccc}
\hline
  & & \multicolumn{2}{c}{Extremely nearby event @ $O$(1 kpc)} 
  & & \multicolumn{2}{c}{Galactic event @ $O$(10 kpc)} 
  & & \multicolumn{2}{c}{Extragalactic event @ $O$(1 Mpc)} \\
  & & \multicolumn{2}{c}{(see Section \ref{sec:near})} 
  & & \multicolumn{2}{c}{(see Section \ref{sec:galactic})} 
  & & \multicolumn{2}{c}{(see Section \ref{sec:extra})}\\
  \cline{3-4} \cline{6-7}  \cline{9-10}
  \multicolumn{2}{c}{signals}
  & detector & horizon && detector & horizon && detector & horizon\\
\hline 
neutrino & pre-SN $\bar{\nu}_{\rm e}$ & KamLand & $< 1$ kpc && --- &&& ---\\
              &                                             & {\it HK (20XX-)} & $<3$ kpc              \\
              & $\bar{\nu}_{\rm e}$ burst 
              & SK & Galaxy$^*$ && SK & Galaxy && {\it HK} & $<$ a few Mpc\\
              & $\bar{\nu}_{\rm e}$ burst
              & {\it JUNO (201X-)} & Galaxy && {\it JUNO} & Galaxy && ---\\
              & ${\nu}_{\rm e}$ burst       
              & {\it DUNE (20XX-)} & Galaxy && {\it DUNE} & Galaxy && ---\\
\hline
GW & Waveform$^\dagger$ & H-L-V-K$^\ddagger$ & $<$ several kpc \\
       & Detection & & && H-L-V-K & $\lesssim 8.5$ kpc && {\it ET (20XX-)} & $\lesssim 100$ kpc\\
\hline
EM & Optical & $< 1$ m class &&& 1--8 m class$^{**}$ &&& $< 1$ m class\\ 
      & NIR        & $< 1$ m class &&& $< 1$ m class &&& $< 1$ m class \\
\hline 
\multicolumn{10}{l}{$^*$Detectable throughout the Galaxy.}\\
\multicolumn{10}{l}{$^{**}$ $\sim 25$\% of SNe are too faint to be detected. (Section \ref{sec:galmm}, see also Figure \ref{fig:opticaldetection})}\\
\multicolumn{10}{l}{$^\dagger$Waveform means detection with sufficient signal-to-noise to unravel the GW waveform.}\\
\multicolumn{10}{l}{$^\ddagger$A network of aLIGO Hanford and Livingston, adVirgo, and KAGRA (Section \ref{sec:detector}).}\\
\end{tabular}
\end{center}
\end{table*}

Already, various aspects of multi-messenger physics of Galactic and nearby CCSNe have been investigated. For example, signal predictions of neutrino and GW messengers have been investigated by many authors. In particular, the first $\sim 500$ milliseconds following core collapse is thought to be critical for a successful explosion, and has been studied with three dimensional hydrodynamics and three-flavor neutrino transport by various authors \citep[e.g.,][]{Kuroda:2012nc,Ott:2012mr,Hanke13,Tamborra:2014hga}. On the other hand, the long-term neutrino emission characteristics have been investigated by several groups based on spherically symmetric general relativistic simulations using spectral three-flavor Boltzmann neutrino transport \citep[e.g.,][]{fischer10}. Similarly, while there are a number of multi-dimensional core-collapse simulations with GW predictions \citep[e.g.,][]{EMuller12,Kuroda14,yakunin15}, detailed detectability studies have been limited \citep[see, however,][]{hayama15,Gossan:2015xda}. 
\citet{leonor10} have discussed utilization of a joint analysis of GW and neutrino data,
particularly for electromagnetically dark (or failed) CCSNe.
The importance of the SBO and their connections to multi-messenger observations have been investigated by, e.g., \cite{Kistler:2012as}. Recently, \cite{adams13} revisited the investigation of dust attenuation of Galactic CCSNe. Using modern models of the Galactic dust distribution, they present the distributions of the observed V-band and near-infrared (NIR) band magnitudes of Galactic CCSNe. They also emphasize the importance of neutrino warning and pointing, as well as the need for a wide-field IR detector for ensuring the detection of the early CCSN light curve. 

In this paper, we revisit the implementations of multi-messenger probes of CCSNe. We improve upon previous studies in several ways. First, we self-consistently determine predictions of neutrinos, GW, and electromagnetic signals from a CCSN based on a long-term two-dimensional axisymmetric simulation (\citealt{nakamura15}; Nakamura et al.~in preparation). To focus on the most common Type IIP supernova, we adopt a non-rotating $17 M_\odot$ red supergiant star with solar metallicity. The simulation gives neutrino and GW signals for the first $\sim 7$ seconds after core bounce. We use the ensuing physical parameters (the radius and mass of the central remnant, the mass of synthesized nickel, and the explosion energy) to estimate the electromagnetic light curve. The time-sequence of multi-messenger signals is summarized in Figure \ref{fig:lumi}. The pre-CCSN neutrino emission of \cite{odrzywolek04}, the neutrino burst and gravitational energy release predicted directly from the numerical simulation, and the analytic bolometric light curve of SBO, plateau, and tail signals are shown and labeled. The neutrino burst luminosity is extrapolated up to 200 seconds based on the gravitational energy release rate from the shrinking protoneutron star. The GW energy emission rate is estimated using a quadrupole formula \citep{finn90,muellerb13}.

In addition, 
we investigate tasks aimed at aiding the prospects of ensuring multi-messenger signal detections of a future CCSN. For this purpose we separately consider CCSN occurring in three distance regimes: 
we consider a CCSN occurring extremely nearby (less than 1 kpc away), 
at the Galactic Center ($\sim 10$ kpc, the most likely distance for a Galactic CCSN), 
and a CCSN in neighboring galaxies (within several Mpc). 
Although neutrino detection can unambiguously indicate a core-collapse event, prompt and precise pointing information is needed for the electromagnetic community to fully exploit the advanced warning. In cases where this fails, either due to the electromagnetic signal appearing too quickly or the neutrino statistics limiting pointing precision, telescopes could resort to searching a precompiled list of potential targets. This is particularly useful in the case of CCSNe in nearby galaxies, whose expected neutrino event rates do not provide a compelling angular pointing. To these ends, we compile a list of known Galactic red supergiant candidates in the vicinity of the Earth, as well as a list of nearby galaxies with estimates of their CCSN rates. 

The detectability of multi-messenger signals are summarized in Table \ref{tbl:summary}. The rows show the multi-messenger signals, 
while the columns show three distance regimes. 
Names in \textit{italic} denote future detectors. 
For neutrinos, only the dominant channels are shown: $\bar{\nu}_e$ for Super-K, Hyper-K, and JUNO, and $\nu_e$ for DUNE. The detectors however have other channels that allow further spectral and flavor information to be extracted \citep[for recent discussions, see, e.g.,][]{Laha:2013hva,Laha:2014yua}. In particular, the neutral current events at liquid scintillator detectors such as JUNO hold the important capability to measure the heavy lepton flavor neutrinos \citep{Beacom:2002hs,Dasgupta:2011wg}. 
One important thing is enhancement of detectability by combining the signals.
For example, we demonstrate that the information of the core bounce timing provided by neutrinos can be used to improve the sensitivity of GW detection. Importantly, this increases the GW horizon from some $\sim 2$ kpc to $\sim 8.5$ kpc (based on our numerical model), which opens up the Galactic Center region to GW detection even for non-rotating progenitors 
(GW signals from collapse of rapidly rotating cores are circularly polarized \citep{hayama16} and significantly stronger \citep[e.g.,][]{Kotake13}). 

The paper is organized as follows. In Section \ref{sec:setup}, we summarize our setup. We describe our core-collapse simulation, methods for calculating multi-messenger signals, and summarize the detectors we consider and the method for determining signal detections. We discuss the case of a CCSN in the Galactic Center in Section \ref{sec:galactic}, the case of an extremely nearby CCSN in Section \ref{sec:near}, and the case of a CCSN in neighboring galaxies in Section \ref{sec:extra}. Sections \ref{sec:galactic}, \ref{sec:near}, and \ref{sec:extra} are all similarly organized in the following way: descriptions of the multi-messenger signals separately, followed by a discussion of the merits and the ideal procedures for their combination. In section \ref{sec:summary}, we conclude with an overall discussion and summary of our results.

\section{Setup}
\label{sec:setup}

In this section we describe the setup of exploring multi-messenger signals from CCSNe. 
We first describe the setup of our numerical CCSN calculation, followed by how neutrino,
GW, and optical signals are calculated. We then discuss multi-messenger
detector considerations. 

\subsection{Supernova model}
\label{sec:models}

The basis of our CCSN model is 
a long-term simulation of an axisymmetric neutrino-driven explosion 
initiated from a non-rotating solar-metallicity progenitor 
of \citet{woosley02} with zero-age main sequence (ZAMS) mass of $17 ~ \Msun$. 
This progenitor model 
has a radius $958 ~ \Rsun$ and mass $13.8 ~ \Msun$ 
at the onset of gravitational collapse of the iron core. 
This progenitor retains its hydrogen envelope and is classified 
as a red supergiant star (RSG). It is therefore expected to explode as Type II supernova.

The numerical code we employ for the core-collapse simulation 
is the same as found in \citet{nakamura15},
except for some minor revisions. 
The spatial range considered in \citet{nakamura15} was 
limited to 5,000 km from the center. For the current study, 
we extend this to 100,000 km in order to study the late-phase
evolution. 
This outer boundary corresponds to the bottom of the helium layer,
and the interior includes $4.1 \, \Msun$ of the progenitor. 
The model is computed on a spherical coordinate grid 
with a resolution of $n_r \times n_\theta = 1008 \times 128$ zones. 

\begin{figure}
\begin{center}
\includegraphics[width=1.0\textwidth, bb=0 0 600 260]{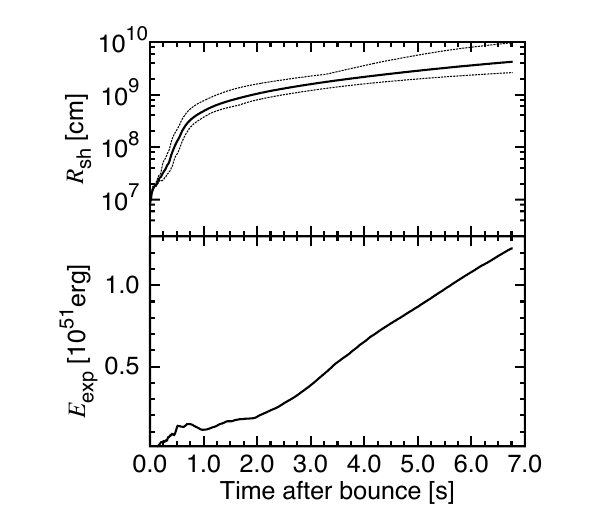} 
\caption{
Time evolution of shock radius (top panel) and explosion energy (bottom panel). 
For shock radius, the maximum, mean, and minimum are shown as separate 
lines from top to bottom. 
}
\label{fig:mshell}
\end{center}
\end{figure}

For electron and anti-electron neutrinos, we use the isotropic diffusion source approximation \citep[IDSA,][]{idsa}, taking 20 energy bins with an upper bound of 300 MeV. 
Regarding heavy-lepton neutrinos, we employ a leakage scheme. 
In the high-density regime, we use the Equation of State (EOS) of \citet{lattimer91} with a nuclear incompressibility of $K = 220$ MeV. 
At low densities, we employ an EOS accounting for photons, electrons, positrons, and ideal gas contribution from silicon.
During our long-term CCSN simulation, we follow explosive nucleosynthesis by solving a simple nuclear network consisting of 13 alpha-nuclei. 
Feedback from the composition change to the EOS is neglected, whereas the energy feedback from the nuclear reactions to the hydrodynamic evolution is taken into account, as in \citet{nakamura14a}. 

Figure \ref{fig:mshell} shows 
time evolution of shock radius, $R_{\rm sh}$, and explosion energy, $E_{\rm exp}$. 
Here the explosion energy is estimated 
by adding up the kinetic, internal, and gravitational energies 
in all zones where the sum of these energies and radial velocity are positive. 
This CCSN model successfully revives its shock at 296 ms after bounce 
and the shock reaches the bottom of the helium layer 
at the end of our simulation at 6.77 s. 
At the termination of the simulation, 
the explosion energy is $1.23 \times 10^{51}$ erg 
and the mass of $^{56}{\rm Ni}$ in unbound material 
(summed up over the zones in the same manner as $E_{\rm exp}$) 
is $0.028 \, \Msun$. 

\begin{figure}
\begin{center}
\includegraphics[width=0.4\textwidth, bb= 90 0 500 500]{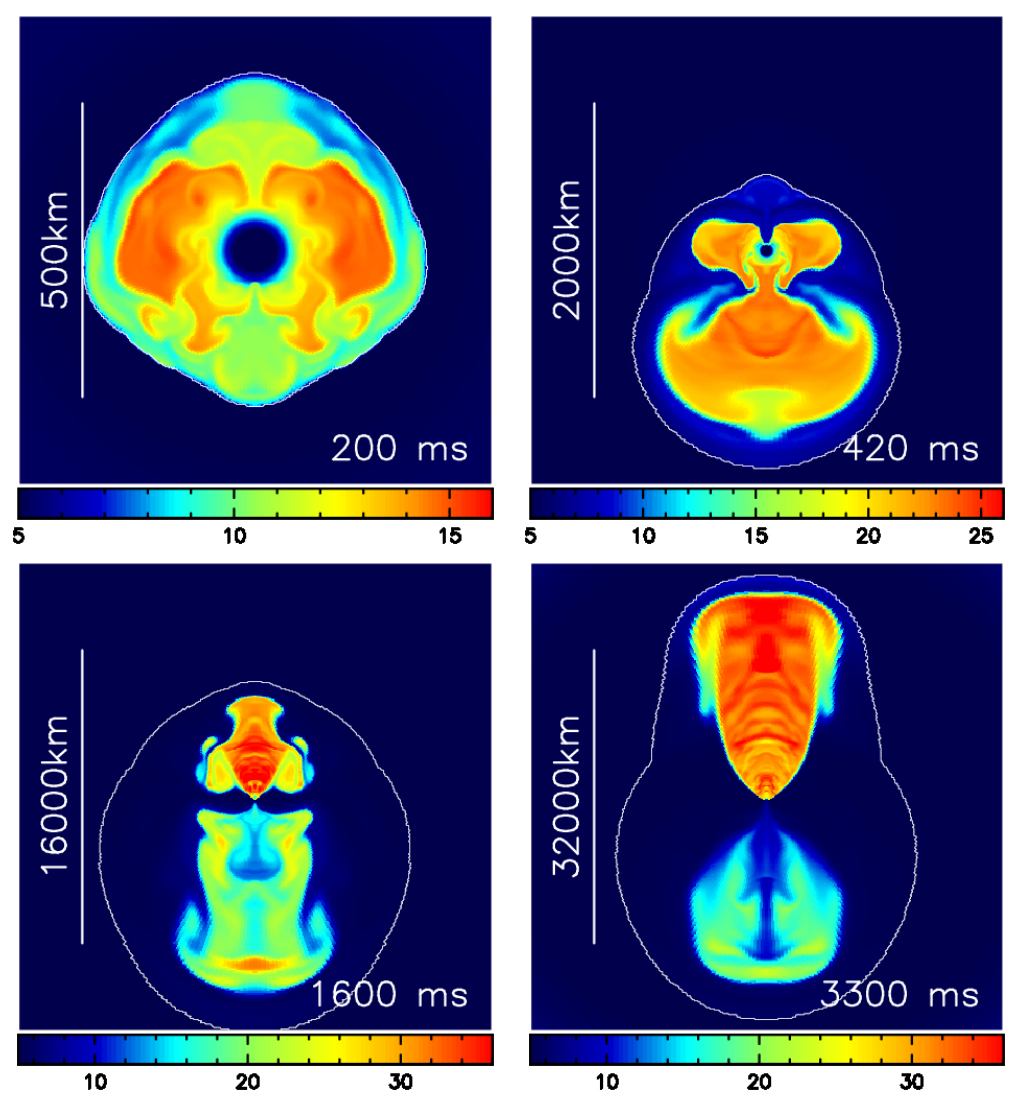} 
\caption{
Entropy distributions in units of $k_{\rm B}$ per baryon
for the CCSN model at four selected postbounce times 
indicated in the bottom right corners. The white contour
represents the shock front. 
Note the different scales in each panel. 
}
\label{fig:snap}
\end{center}
\end{figure}

Figure \ref{fig:snap} visualizes the entropy distribution at four time snap shots. 
The formation of hot bubbles (the regions colored by red in top left panel
of Figure \ref{fig:snap}) is clear evidence of neutrino-driven convection,
 which revives the stalled bounce shock by the (so-called)
 neutrino-driven mechanism \citep{bethe90}. In our model,
the runaway expansion of the stalled shock initiates at about 200 ms
postbounce. In Figure \ref{fig:snap}, the shock front (white contours) exhibits 
a dipolar deformation along the poles, as previously identified in two-dimensional (2D) simulations
\citep{bruenn13,summa15}, with the average shock radii
becoming bigger with time (from top left to bottom right panel).
After the shock passes the carbon core, the shape of the bipolar 
explosion (occurring stronger towards the north pole
than the south pole, as seen in the bottom right panel of Figure \ref{fig:snap}) is kept nearly
 unchanged, and the shock expands rather in a self-similar fashion
 from thereon.

\subsection{Limitations of our model}
\label{sec:limit}

Throughout this paper we use the CCSN model described in the above section. 
The properties of CCSN model, however, depend on the structure of progenitor stars as well as input physics in simulations. Neutrino luminosity, for example, changes for different mass (or ``compactness'') of progenitors and different choice of EOS \citep{oconnor13,nakamura15}. 
Our simulations for 101 progenitor models with solar metallicity, employing the same numerical scheme and EOS as the current study, show diverse neutrino luminosities; $1.0$ -- $3.5 \times 10^{52}$ erg s$^{-1}$ at 0.5 s for anti-electron neutrino, $17 ~ \Msun$ model lying in between ($1.7 \times 10^{52}$ erg s$^{-1}$).
Note that the mass of the progenitor ($17 ~ \Msun$) is not extreme and the EOS we use is widely accepted to CCSN modelers.

Numerical results also depend on unphysical variables, such as spatial resolution of simulations. We have tested the effects of angular resolution of spatial grid on CCSN properties. Besides a model with fiducial resolution of 128 angular zones, we used three additional models having 192, 256, and 320 angular zones. The numerical setup of the models was identical to the description in section \ref{sec:models}. For these models, differences in the shock dynamics as well as CCSN properties like explosion energy were observed as expected. Explosion energy, gravitational mass of PNS, and nickel mass in unbound material, estimated at 5.5 s after bounce, range in 
$E_{\rm exp} = 0.42$ -- $0.98 \times 10^{51}$ erg, 
$M_{\rm PNS} = 1.78$ --$1.91 \, \Msun$, and 
$M_{\rm Ni} = 0.013$ -- $0.025 \, \Msun$, respectively. 
Our resolution study does not show any converging results and the resolution changes do not produce uniform trend, which is consistent with recent 2D results by \citet{summa15}. 
It implies the strong influence of stochastic turbulent motions behind the shock front and difficulty in predicting any definitive CCSN properties given a progenitor structure. 

What we present in this paper is just one example of simulated CCSNe, but it is quite important to discuss multi-messenger signals focusing on one concrete example modeled by the state-of-the-art calculation.

\subsection{Multi-messenger emissions}
\label{sec:multimessengeremission}

We study three kind of multi-messenger signals from the CCSN: neutrinos, GWs, and electromagnetic waves. Below, we briefly summarize the modeling of each signal, keeping the details in Appendix \ref{sec:appsignal}. 

{\it Neutrinos}: 
Neutrino signals are a direct outcome of our CCSN simulation which takes account neutrino transport in a self-consistent way. After emission from the collapsed core, neutrinos undergo flavor conversions during propagation to the Earth. In general, neutrino oscillation varies during the core-collapse evolution, and also depends on the neutrino mass hierarchy. Additional flavor mixing is induced by the coherent neutrino-neutrino forward scattering potential \citep[for a review, see, e.g.,][]{Duan:2010bg,Mirizzi:2015eza}, and exact predictions of collective flavor transformation in the post-bounce environment, and when it will apply, and at what energies, are still under debate. 
Therefore, we consider as a simple scenario the Mikheyev-Smirnov-Wolfenstein (MSW) matter effects in the normal and inverted mass hierarchy, as well as a no mixing and a full flavor swap scenario. While these scenarios are not equality realistic, they are aimed to cover the extremes (see Appendix \ref{sec:appnu} for details).

{\it Gravitational waves}: 
We extract the GW signal from our simulations using the standard quadrupole formula \citep[e.g.,][]{Misner:1974qy}. For numerical convenience, we employ the so-called first-moment-of-momentum-divergence (FMD) formalism proposed by \cite{finn90} (see also modifications in \citet{Murphy:2009dx,muellerb13}). The angle between the symmetry axis in our 2D simulation and the line of sight of the observer is taken to be $\pi/2$ to maximize the GW amplitude (see Appendix \ref{sec:appgw} for details). 

{\it Electromagnetic signals}: 
The first electromagnetic signal from the CCSN is the emission from SBO, followed by the characteristic plateau phase for Type IIP supernovae. Given the large difference in time scales involved in our CCSN model ($\sim 7$ s) and the emergence of the SBO signal (a few minutes to a day), it is difficult for the current simulations of core collapse to directly reproduce the electromagnetic signals. Therefore, we use the final physical parameters of our simulation and employ analytic expressions for the luminosity and duration in SBO \citep{matzner99}, and in the plateau phase \citep{popov93} (see Appendix \ref{sec:appem} for details).

\subsection{Detector considerations}
\label{sec:detector}

Below, we summarize the detectors setup assumed in this paper. 
In this paper, we will consider CCSNe
at three distance regimes: 
occurring at the Galactic Center (section \ref{sec:galactic}), 
a very nearby location (section \ref{sec:near}), 
and extragalactic distances (section \ref{sec:extra}).
As we discuss in the subsequent sections, the necessary detector depends 
strongly on these distance regimes.

{\it Neutrinos}: 
We consider two water \v{C}erenkov neutrino detectors, Super-K\footnote{http://www-sk.icrr.u-tokyo.ac.jp/sk/index-e.html} and its successor Hyper-K \citep{Abe:2011ts}. Super-K has a fiducial volume of 32.5 kton in neutrino burst mode with a low energy threshold of 3 MeV. For  Hyper-K, we assume a burst-mode fiducial volume of 740 kton and an energy threshold of 7 MeV \citep{Abe:2011ts}. The main detection channel is of $\bar{\nu}_e$ with inverse-beta decay (IBD), which provides good kinematic but poor angular information of the incoming neutrinos \citep{Vogel:1999zy}. We thus also consider electron scattering, which is sensitive to all neutrino flavors and is forward peaked \citep{Vogel:1989iv}. 
Details are included in Appendix \ref{sec:appnu}.

{\it Gravitational waves}: 
We consider a network of interferometric GW detectors consisting of four detectors: aLIGO\footnote{https://www.advancedligo.mit.edu} Hanford and Livingston, adVirgo\footnote{http://wwwcascina.virgo.infn.it/advirgo}, and KAGRA\footnote{http://gwcenter.icrr.u-tokyo.ac.jp/en} at their geophysical locations (H-L-V-K). For each detector the noise spectrum is taken from 
\citet{ligodesign}, \citet{advv}, and \citet{aso13}, 
and we assume sensitivities that are expected to be realized by the year 2018. The detection of the gravitational waveform from a CCSN is determined using this global network of GW detectors by the coherent network analysis \citep{Klimenko:2005,2010NJPh...12e3034S, hayama07, hayama15}. The coherent network analysis was introduced by \citet{1989PhRvD..40.3884G}, the basis of which is to reconstruct arbitrary gravitational waveforms by linear combination of data from GW detectors. For the signal detection, reconstruction, and source identification, we perform Monte Carlo simulations using the {\tt RIDGE} pipeline \cite[see][for more details]{hayama07}. 

{\it Electromagnetic signals}:
We consider electromagnetic observations
in optical and NIR wavelengths.
For detection of the electromagnetic signals,
interstellar extinction proves to be critical \citep{adams13}. 
Since NIR wavelengths are only mildly affected by the extinction,
Galactic CCSNe can be detected with small-size NIR telescopes.
On the other hand, the optical brightness is 
significantly attenuated by dust extinction,
which depends on the location of the CCSN event.
In addition, since 
the positional localizations by GW and neutrino detections are typically larger 
than a few degree,
wide-field capability is also 
crucial not to miss the very first electromagnetic signals.
Therefore, for optical telescopes,
we consider various sizes of wide-field telescopes/facilities,
such as the All-Sky Automated Survey for SuperNovae 
\citep[ASAS-SN,][]{shappee14}, Evryscope \citep{law15},
Palomar Transient Factory \citep[PTF,][]{law09,rau09},
Pan-STARRS1 \citep[PS1, e.g.,][]{kaiser10},
Subaru/Hyper Suprime-Cam \citep[HSC,][]{miyazaki06,miyazaki12},
and Large Synoptic Survey Telescope \citep[LSST,][]{ivezic08}.

\section{Galactic Center Supernovae}
\label{sec:galactic}

\subsection{Neutrino signal}
\label{sec:galnu}

A Galactic CCSN, occurring at a distance of $8.5$ kpc, will provide a rich trove of neutrino data in present and future neutrino detectors \citep[for a recent review, see, e.g.,][]{Mirizzi:2015eza}. The observed event rate in the detector is,
\begin{equation}
\frac{dN_e}{dT_e} = N_t \int^{\infty}_{E_{\rm min}} d E_\nu \frac{dF_\nu}{dE_\nu}(E_\nu) \frac{d \sigma}{dT_e}(E_\nu,T_e),
\end{equation}
where $N_t$ is the number of appropriate targets, $d\sigma/dT_e(E_\nu,T_e)$ is the differential cross section for the appropriate channel, and $dF_\nu/dE_\nu (E_\nu)$ is the neutrino flux observed at the Earth for the appropriate flavor. The threshold neutrino energy $E_{\rm min}$ is determined by the threshold energy of a lepton that can be detected. The neutrino flux observed at the Earth is,
\begin{equation}
\frac{dF_\nu}{dE_\nu}(E_\nu) = \frac{L_\nu}{4 \pi D^2 \langle E_\nu \rangle } f(E_\nu),
\end{equation}
where $L_\nu$ is the neutrino luminosity, $\langle E_\nu \rangle$ is the mean neutrino energy, $D$ is the distance to the CCSN,  and $f(E_\nu)$ is the neutrino spectral distribution. For the latter we adopt the pinched Fermi-Dirac spectrum \citep{Keil:2002in},
\begin{equation}
f(E_\nu) = \frac{(1+\alpha)^{(1+\alpha)}   }{ \Gamma(1+\alpha) }  \frac{E_\nu^\alpha}{\langle E_\nu \rangle^{\alpha+1}} e^{-(1+\alpha) \frac{E_\nu} { \langle E_\nu \rangle}},
\end{equation}
where $\alpha$ describes the pinching
given by $\alpha = (\varepsilon_2 - 2 \varepsilon_1^2) / (\varepsilon_1^2 - \varepsilon_2)$ 
using $\varepsilon_n = \int^\infty_0 \varepsilon^n f(\varepsilon) \, d\varepsilon$; 
$\alpha = 2.3$ corresponds to a Fermi-Dirac distribution. This spectral form is a good fit to the neutrino spectra of CCSN simulations \citep[e.g.,][]{Tamborra:2012ac}. 

\begin{figure}
\begin{center}
\includegraphics[width=0.2\textwidth, bb=200 50 500 570]{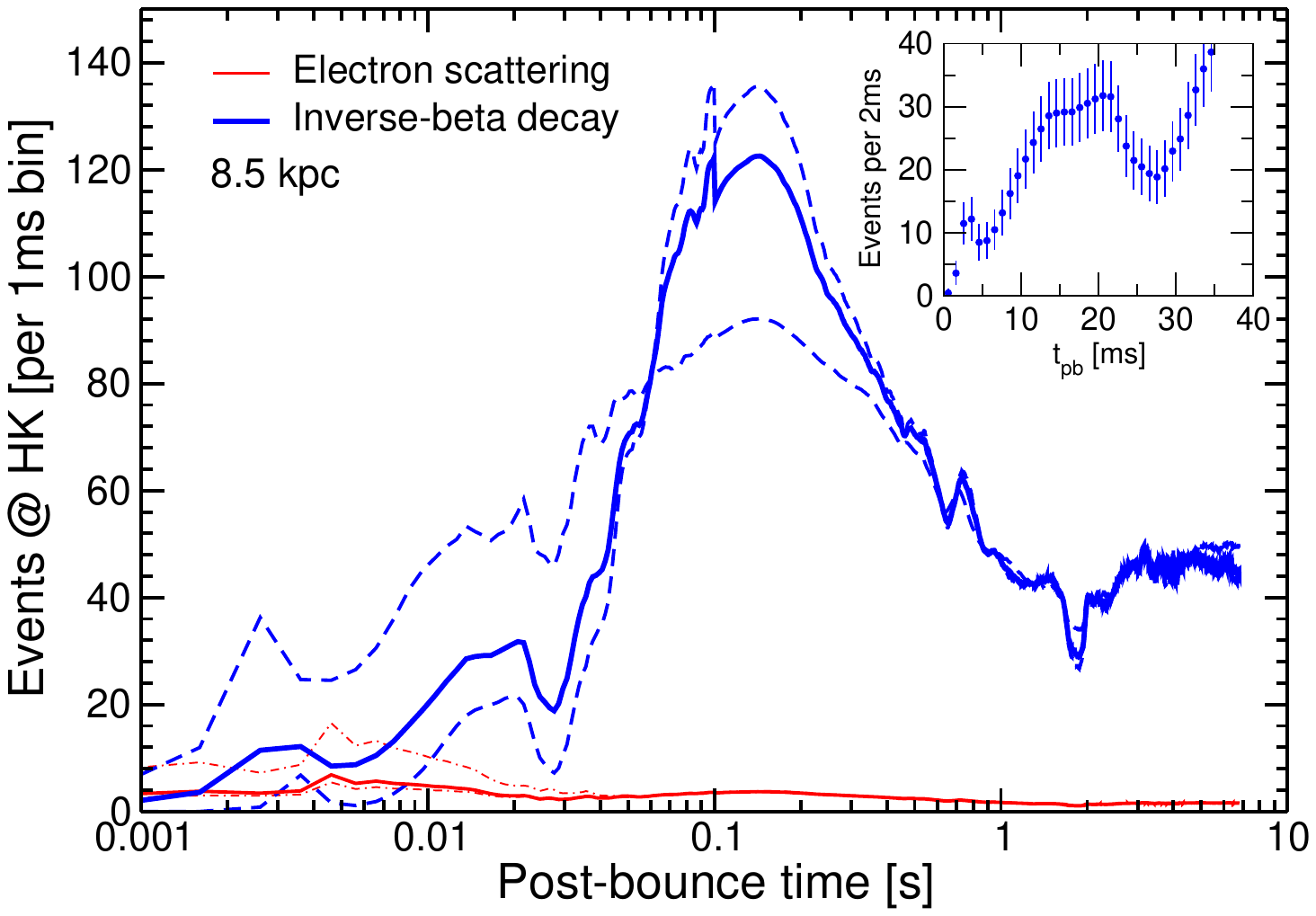} 
\caption{Expected neutrino events at Hyper-K in 1ms time bins, for a Galactic CCSN at a distance of 8.5 kpc. 
Inverse-beta events are shown by thick blue lines, while the lower thin red lines denote electron scattering events. For each, the three lines show three neutrino mixing scenarios: (i) MSW mixing (central solid lines), (ii) no mixing (upper lines), and (iii) full flavor swap (lower lines). This is done to show the range of possible mixing, accounting for both mass hierarchies, MSW, and collective neutrino oscillations. The MSW mixing line represents normal (inverted) mass hierarchy for $\bar{\nu}_e$ ($\nu_e$). The no mixing assumes the observed electron neutrinos are entirely composed of the same flavor at source (i.e., $\bar{\nu}_e = \bar{\nu}_e^0$ and ${\nu}_e = {\nu}_e^0$), while the full swap assumes they are entirely exchanged with the heavy flavor states at source (i.e., $\bar{\nu}_e = {\nu}_x^0$ and ${\nu}_e = {\nu}_x^0$); see appendix \ref{sec:appnu} for details. 
The inset shows the first 40 ms of the IBD signal at Hyper-K, shown in 2 ms bins, for the MSW mixing case. The error bars are root-N Poisson errors.
}
\label{fig:neutrino}
\end{center}
\end{figure}

We obtain the values of $(L_\nu,\langle E_\nu \rangle, \alpha)$ for $\nu = (\nu_e, \bar{\nu}_e, \nu_x)$ from our core-collapse simulation\footnote{Except $\alpha$ for $\nu_x$. The leakage scheme is adopted for $\nu_x$ and spectrum information is not available. Here we fix $\alpha = 2.3$ for $\nu_x$.}, 
and show the event rate per 1 ms bin expected for a 8.5 kpc event at the Hyper-K detector in Figure \ref{fig:neutrino}. The thick blue and thin red lines denote inverse beta and $e^-$-scattering events, respectively. The central solid lines denote MSW mixing, while the upper and lower lines denotes the no-mixing and full-mixing cases, respectively. The time-integrated number of events is 328,000--329,000 inverse-beta events and 11,300--11,700 $e^-$ scattering events.

\begin{figure*}
\begin{center}
\begin{tabular}{cc}
\includegraphics[width=0.1\textwidth, bb=770 0 970 800]{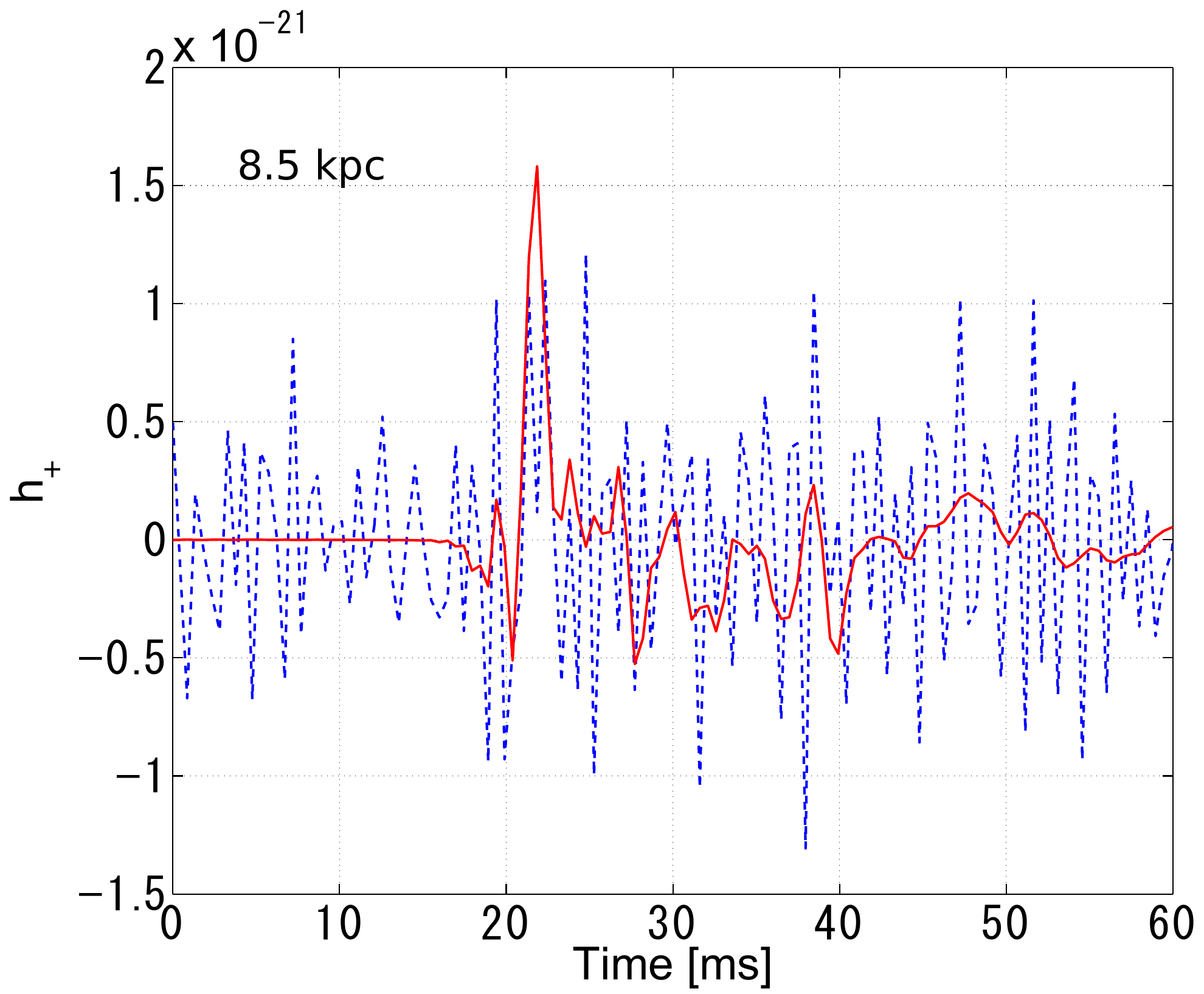} &
\includegraphics[width=0.1\textwidth, bb=0 0 200 800]{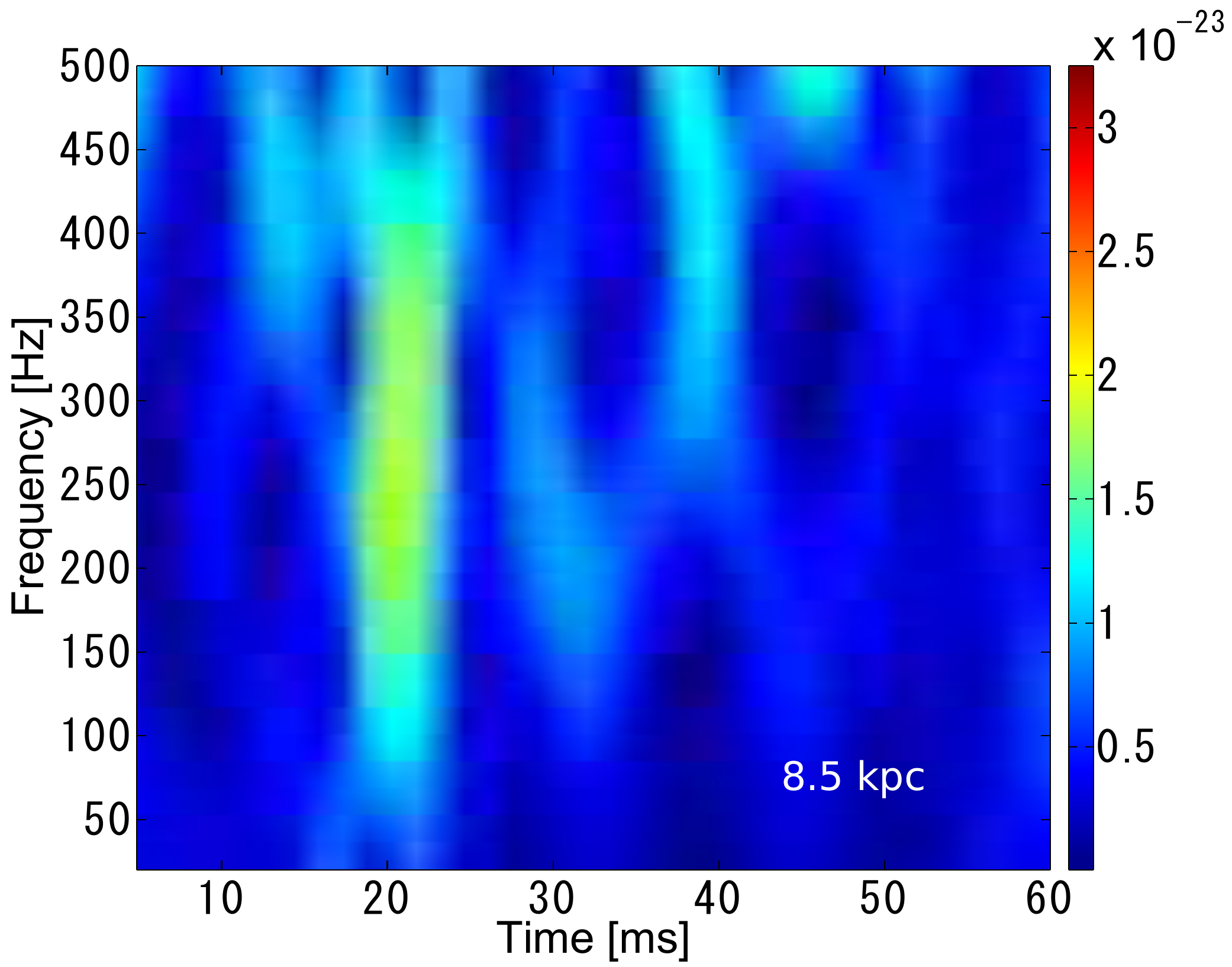}
\end{tabular}
\caption{
The GW characteristics in the first 60 ms postbounce. 
Left: the inputted (solid red line) and reconstructed (dashed blue) gravitational waveform. 
Right: the spectrogram of the reconstructed waveform in the frequency window [50, 500] Hz. 
Both panels are for a CCSN at a distance of 8.5 kpc.
} 
\label{fig:tfgc}
\end{center}
\end{figure*}

\begin{figure*}
\begin{center}
\begin{tabular}{cc}
\includegraphics[width=0.1\textwidth, bb=770 0 970 730]{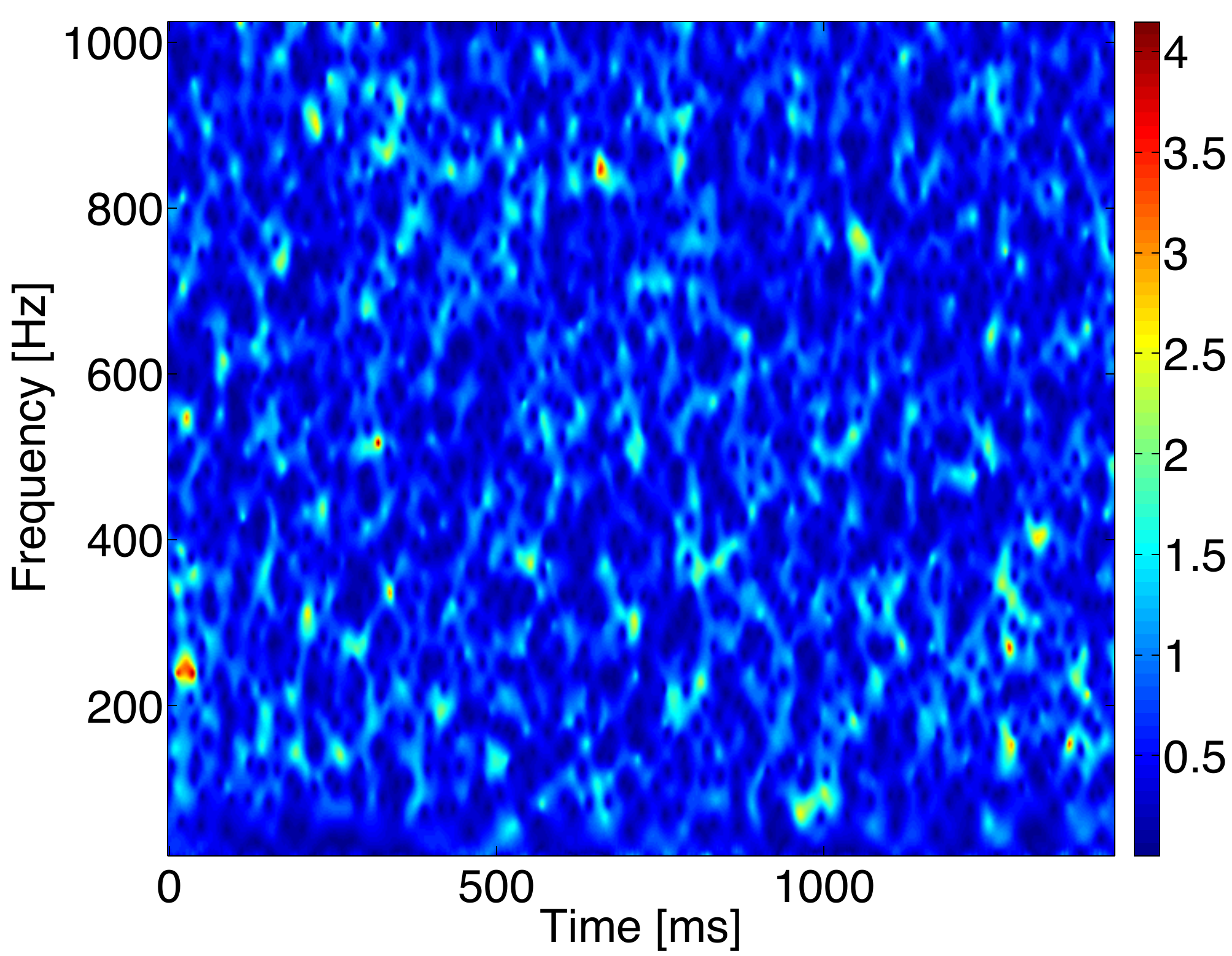} &
\includegraphics[width=0.1\textwidth, bb=0 0 200 730]{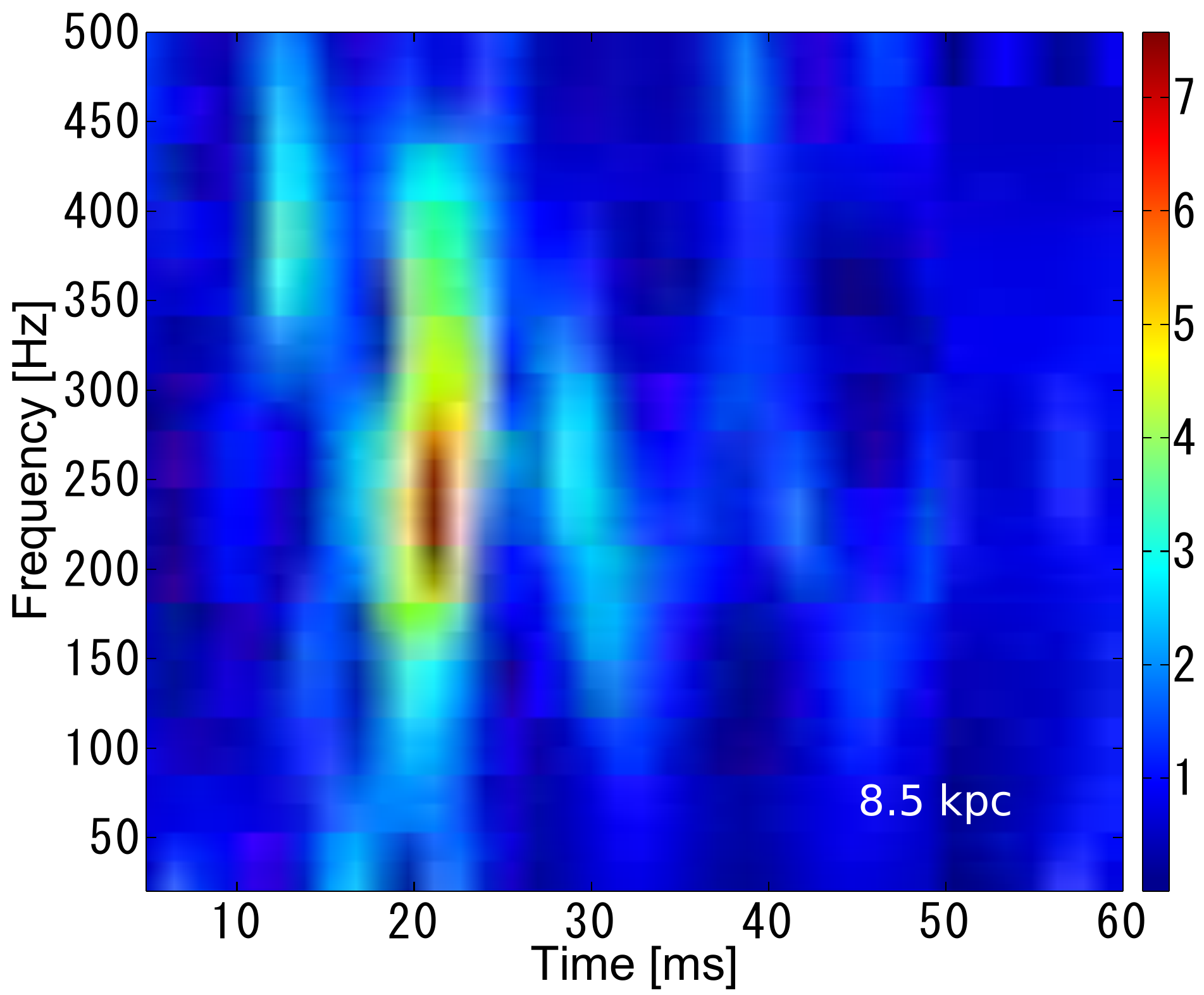}
\end{tabular}
\caption{
SNR of the GW from a distance of 8.5 kpc estimated in time-frequency pixels. 
Left: analysis based on a GW search over more than 1 second without a neutrino trigger.
Right: SNR in the small time-frequency window with the aid of the neutrino timing information, corresponding to the right panel of Figure \ref{fig:tfgc}. 
Note the different scale between the left and right panels. 
} 
\label{fig:gcsnr}
\end{center}
\end{figure*}

The high-statistics light curve will enable various studies, such as probes of the standing accretion shock instability \citep[SASI; see, e.g.,][for a recent review]{Foglizzo:2015dma} that affect the luminosities and energies of the emitted neutrinos. The detectability of SASI has been explored by various studies focusing on both Hyper-K and IceCube for Galactic CCSNe 
\citep{Lund:2010kh,Lund:2012vm,Tamborra:2013laa,Tamborra:2014hga}. Our particular progenitor explodes predominantly by neutrino-driven convection and hence the SASI modulated signal is weak. The neutronization burst will also allow unique probes of the core collapse \citep{Kachelriess:2004ds}. During this phase, the collapse is spherical enough that the bulb model, i.e., neutrinos emitted from a sharp spherical surface, is a reasonable approximation for neutrino mixing calculations \citep[e.g.,][]{Cherry:2012zw}. 

The time of core bounce can be estimated from the detected neutrinos. The inset of Figure \ref{fig:neutrino} shows the initial 40 ms of the IBD signal, where the error bars show the root-${N}$ statistical error. The statistics will be sufficiently high that the bounce time can be estimated with high accuracy. The IceCube detector will similarly have high significance detection of the $\bar{\nu}_e$ light curve through their IBD on free protons in ice \citep{Dighe:2003be,Kowarik:2009qr}. While individual neutrino events cannot be reconstructed at IceCube, a statistically significant ``glow'' is predicted during the passage of the supernova neutrino burst. For a Galactic supernova at 8.5 kpc, the bounce time is estimated to be measurable to within $\pm 3.0$ ms at 95\% confidence level \citep{Halzen:2009sm}. 

Neutrinos also provide pointing information. Using forward $e^-$ scattering events, a galactic core-collapse event can be pointed to within an error circle of some $6^\circ$ with the current Super-K detector \citep{Beacom:1998fj}. The main background is the near isotropic IBD signal. 
Coincidence tagging is a powerful way of distinguishing IBD from other events and backgrounds. With the current setup using pure water, the neutrons from IBD capture on protons yielding a 2.2 MeV gamma which falls below the lowest energy trigger threshold at Super-K. Using a forced trigger, tagging efficiencies of $\sim 20$\% have been obtained for diffuse supernova neutrino searches \cite{Zhang:2013tua}. Upon completion of its Gadolinium upgrade, Super-K is expected to be able to neutron-tag the IBD with $\sim 90$\% efficiency \citep{Beacom:2003nk}. The increased tagging efficiency will improve the pointing accuracy to $\sim 3^\circ$ at Super-K, and if Hyper-K similarly has 90\% tagging efficiency, to some $0.6^\circ$ \citep{Tomas:2003xn}. 
The DUNE detector is expected to provide comparable sensitivity. Scaling the results of \cite{Bueno:2003ei} to the 34 kton detector volume of DUNE and to our adopted 8.5 kpc distance, the expected uncertainty in supernova pointing is $\sim 1.5^\circ (34 {\rm \, kton} / 1.2 {\rm \, kton})^{-1/2} \sim 0.3^\circ$. 

As will be discussed below, the timing and pointing accuracy directly impacts the scientific benefits for multi-messenger studies of CCSNe. 

\subsection{Gravitational waves}
\label{sec:galgw}

In this section we discuss the detection prospects of GW signals 
from a CCSN occurring at a position close to the Galactic Center. 
We assume the GW source position at right ascension 17.76 hours and declination -27.07 degrees,
and similar to the previous section a distance from the Earth of 8.5 kpc. 
The event time is set to be UTC 2013-02-22 12:12:02, but it is not important for detection efficiency of the detectors under consideration \citep{hayama15}.
The gravitational waveform we employ as an input in this work is
consistent with that obtained in previous GW studies \citep{Emuller04,muellerb13,yakunin15} 
based on self-consistent 2D models.
The overall waveform (top panel of Figure \ref{fig:reconBetel1}) is characterized by
a sharp rise shortly after bounce ($\lesssim$ 50 ms postbounce) due
to prompt convection, which is followed by spikes with increasing
GW amplitudes (maximally $6 \times 10^{-20}$ in Figure \ref{fig:reconBetel1})
in the non-linear phase due to neutrino-driven convection
(and SASI) ($\lesssim$ 600 ms postbounce for this model). Later on,
as the neutrino-driven wind phase sets in (typically $\gtrsim 1$ s
postbounce), and the amplitude shows a gradual decrease.

\begin{figure*}
\begin{center}
\begin{tabular}{cc}
\includegraphics[width=0.48\linewidth, bb=0 0 200 150]{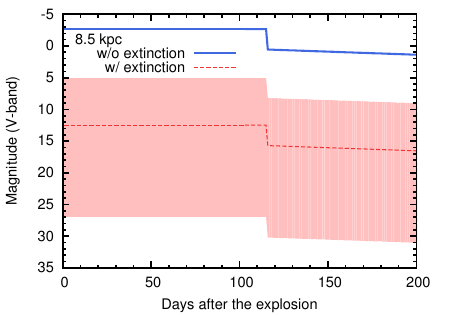} &
\includegraphics[width=0.48\linewidth, bb=0 0 200 150]{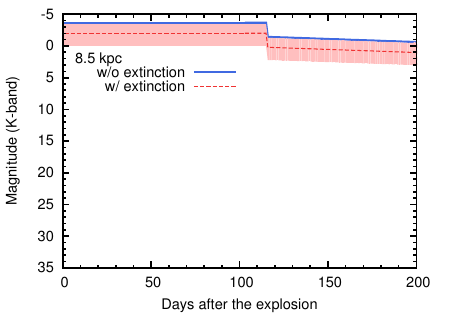} 
\end{tabular}
\end{center}
\caption{
Optical (left panel) and NIR (right panel) signals from a Galactic CCSN at a distance of 8.5 kpc. The solid (blue) lines show schematic light curves of the plateau and tail phases based on the bolometric luminosity models given by \citet{popov93} and \citet{Nadyozhin94}. The expected light curves taking dust extinction into account are shown by dashed (red) lines with a hatched (pink) regions covering the range of half-maximum probability. The extinction correction significantly reduces the optical brightness, whereas the NIR brightness is less affected. The spatial distribution of the apparent magnitudes are shown in Figure \ref{fig:appmag}. Note that the optical and NIR magnitudes of the SBO emission are expected to be fainter than plateau magnitudes by about 1 mag and 2 mag, respectively \citep{tominaga11}.
}
\label{fig:OptNIR}
\end{figure*}

Since we make coordinated observation of the CCSN 
with Super-K and GW detectors, 
the time of the core bounce can be estimated 
from neutrino observations. 
This enables us to restrict 
the time window, e.g., to $[0, 60]$ ms from the estimated time of the core bounce, 
as well as the frequency window to be $[50, 500]$ Hz 
which is suggested to be roughly in the peak frequency range of 
prompt convection \cite[e.g.,][]{muellerb13}.

We apply the gravitational waveform data, superposed on the noise signals, 
within the time--frequency window of $[0, 60]$ ms--$[50, 500]$ Hz 
to the analysis pipeline 
and obtain a reconstructed gravitational waveform.
The left panel of Figure \ref{fig:tfgc} 
compares the inputted original gravitational waveform (solid red line) 
with the reconstructed one (dashed blue line), 
and the right panel shows the spectrogram 
of the reconstructed gravitational waveform. 
In this time--frequency window
the noise dominates the reconstructed waveform 
and it is hard to see any time-dependent waveform structure. 
In the spectrogram, however, 
a highlighted region appears at $t \sim 20$ ms
which originates from the prompt convection.

This feature is observable. Figure \ref{fig:gcsnr} shows the signal-to-noise ratio (SNR) of time--frequency pixels defined by
\begin{equation}
{\rm SNR} = \int \frac{\tilde{x}(f,t)}{S_{\mathrm{n}}(f,t)} \, \mathrm{d}t \, \mathrm{d}f,
\end{equation}
where $\tilde{x}(f,t)$, and $S_{\mathrm{n}}(f,t)$, denotes time--frequency pixels, and onesided-spectrogram densities, at a given frequency and time, respectively. The left panel shows the result without a neutrino trigger, i.e., a GW search over more than 1 second, while the right panel shows the result considering timing information from neutrino observations. The maximal SNR for the prompt convection GW signal pixel increases from $\sim 3.5$ to $\sim 7.5$. The latter almost meets the conventional detection threshold. 

\subsection{Electromagnetic waves}
\label{sec:galem}

The first electromagnetic signal from a CCSN is the emission
from SBO \citep[e.g.,][]{falk78,klein78,matzner99}.
The effective temperature of the SBO emission is
estimated to be $\sim 4 \times 10^5 $K.
Thus, the emission peaks at UV wavelengths.
However, as discussed below, CCSNe at the Galactic Center are likely to
suffer from large interstellar extinction.
Therefore, the observed 
spectral distribution of the SBO
is likely not to peak at UV wavelengths, and observations in
optical and NIR are more promising \citep{adams13}.
For Type IIP supernovae, the SBO emission in optical and NIR wavelengths 
is expected to be fainter than the main plateau emission, which we discuss below,
by about 1 mag and 2 mag, respectively \citep{tominaga11}.

After cooling envelope emission following shock breakout emission
\citep[e.g.,][]{chevalier08,nakar10,rabinak11},
Type IIP supernovae enter the plateau phase lasting about 100 days.
The luminosity and duration of the plateau can be estimated 
by equations (\ref{eq:Lplt})--(\ref{eq:tplt}) using $\Mej$, $E_k$, and $R_0$.
The solid (blue) lines in Figure \ref{fig:OptNIR} 
show schematic light curves
after the plateau phase for our s17.0 model placed at 8.5 kpc distance.
The luminosity is then converted to 
optical ($V$-band, 0.55 $\mu$m) and NIR ($K$-band, 2.2 $\mu$m) magnitudes
assuming a bolometric correction of $M_{\rm bol} - M_V \simeq 0$
and typical color of $M_V - M_K \simeq 1$ \citep{bersten09}.

After the plateau phase, supernovae enter the radioactive tail phase,
which is powered by the decay of \Cofs\
(daughter of \Nifs\ synthesized at the explosion).
For the bolometric luminosity at this phase, 
we assume the radioactive decay luminosity
with an ejected \Nifs\ mass of 0.028 $\Msun$.
Then the luminosity is translated to optical and NIR magnitudes
assuming $M_{\rm bol} - M_V \simeq 0$ and $M_V - M_K \simeq 2$ \citep{bersten09}.

\begin{figure*}
\begin{center}
\begin{tabular}{cc}
  \includegraphics[scale=1.1, bb=0 0 500 200]{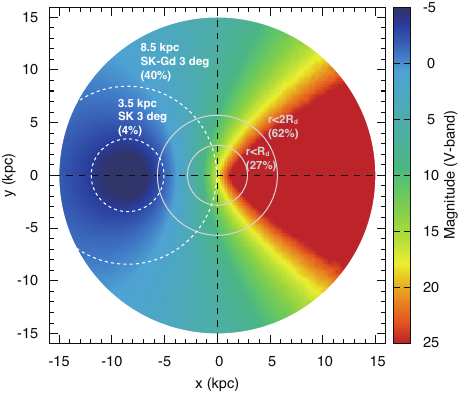} &
  \includegraphics[scale=1.1, bb=270 0 800 200]{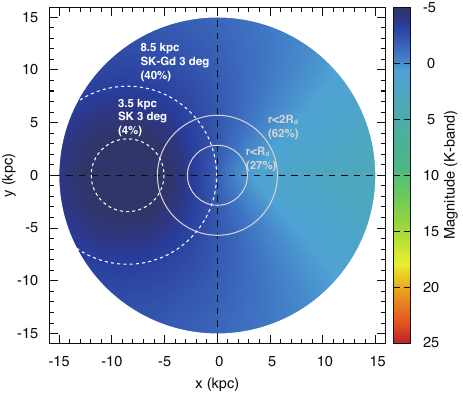}
\end{tabular}
  \caption{Top-down view of the Milky Way galaxy, colored by the expected CCSN plateau optical magnitude (left panel) and NIR magnitude (right). The SBO emission is fainter by 1--2 mag than shown. The Earth is positioned at $(x,y)=(-8.5,0)$ kpc. The NIR magnitude is always bright enough ($\lesssim$ 0 mag) irrespective of the location of the CCSN. On the other hand, the optical magnitude has a wide range depending on the location. Especially, the optical brightness of a CCSN at the opposite end of the Galaxy is significantly affected by dust extinction since the line-of-sight to the CCSN traverses more dusty regions (orange and red colors). The white dashed and grey dashed circles represent constant distances from the Earth and the Galactic Center, respectively. The circles are labeled by their radii, as well as the percentages of the Galactic CCSN rate that they contain. For circles centered on the Earth, the pointing accuracy that can be achieved by Super-K (SK) or Super-K with Gadolinium (SK-Gd) for a CCSN at the circumference is also labeled. 
}
\label{fig:appmag}
\end{center}
\end{figure*}

Despite these crude assumptions,
Figure \ref{fig:OptNIR} still captures the important properties of
optical and NIR emissions
of Galactic CCSNe. 
Both optical and NIR emissions
are extremely bright, $< -2 $ mag,
if we do not take into account interstellar extinction (thick blue lines).
However, since we have more probability to have CCSNe near the center of the Galaxy,
the observed brightness will be significantly affected by interstellar extinction.

Here we estimate the typical range of extinction as follows, 
in a similar way as was done by \citet{adams13}. 
First, we assume a simple three-dimensional Galactic model
both for the CCSN rate and dust distribution:
$\rho (R,Z) \propto e^{-R/R_d}e^{-z/H}$,
where $R$ is the Galactocentric radius,
$z$ is the height above the Galactic plane, and
$R_d$ and $H$ are the scale length
and scale height of the Galactic disk, respectively.
We adopt $R_d = 2.9$ kpc and $H = 110$ pc as in \citet{adams13}.
The Sun is placed at $R=8.5$ kpc and $z = 24$ pc.
The normalization of the dust distribution is determined such that
the extinction toward the Galactic Center is $A_V = 30$ mag.

The expected brightness maps under these assumptions
are shown in Figure \ref{fig:appmag}. The optical brightness 
has a large diversity depending on the
position (left panel): it is brighter than 5 mag for nearby events while
fainter than 25 mag behind the Galactic Center.
On the other hand, the NIR brightness is brighter than
5 mag in almost all locations, as was also demonstrated by \citet{adams13}.
Note that Figure \ref{fig:appmag} can also be used as the brightness distribution of
the SBO emission if optical and NIR magnitudes are shifted down
by about 1 mag and 2 mag, respectively.

The expected observed light curves for CCSN events at $R<R_d$,
which corresponds to about 27 \%\ of Galactic CCSNe,
are shown in Figure \ref{fig:OptNIR}.
The dashed (red) line shows the brightness of the most probable case
and the hatched region covers the range of half-maximum probability.
As shown in the figure, the optical brightness is significantly
affected by the extinction.

In contrast to the optical brightness, the
NIR brightness is less significantly affected by extinction
because extinction is much smaller in NIR, $A_K \sim 0.1 A_V$.
Even in the worst case, the brightness at the plateau phase
is as bright as 0-1 mag.
This can be observed with the preparation of NIR telescopes/instruments
usable for very bright objects (see discussions by \citealt{adams13}).

\subsection{Multi-messenger observing strategy}
\label{sec:galmm}

As emphasized in \cite{adams13}, neutrinos provide crucial triggers to address important questions of \textit{IF} astronomers should look for a Milky Way CCSN, \textit{WHEN} they should look, and \textit{WHERE} they should look. Using our CCSN model based on a long-term numerical simulation, we follow the time-series of events following a core collapse. We emphasize the new insights gained when GWs are included in the discussion. 

To alert the community \textit{IF} one should look for a Galactic supernova, an alarm system called the SuperNova Early Warning System (SNEWS) is in place \citep{Antonioli:2004zb}. The system consists of a network of neutrino detectors that triggers an alert when multiple detectors in different geographical locations report a burst within a certain period. The system is rapid as it does not require human intervention. 
The Super-K detector is large enough that a core collapse in the Milky Way will yield a high statistics neutrino burst detection \citep{abe16}. 
EGADS (Evaluating Gadolinium's Action on Detector Systems), is situated next to Super-K and is planned to be developed into a neutrino burst monitor that reports a burst within the one-minute mark while minimizing both false positives and false negatives \citep{vagins12,adams13}.

Acquiring accurate timing information is important for improving the sensitivity for the detection of GWs. Since the coherent network analysis does not assume a priori information about a waveform, data which do not contain GWs cause degradations for detection. The accurate timing information of core bounce will thus enable one to avoid such data. The neutrinos provide timing information with $1\sigma$ uncertainties of $\sim 10$ ms, and therefore the GW search can be started $10$ ms before the time indicated by the neutrino observation (hereafter referred to as $t_n$). The first strong amplitude of gravitational waveforms from a CCSN is known from numerous simulations to be driven by prompt-convection, and peaks within some $60$ ms after core bounce \citep[e.g.,][]{muellerb13}. The GW analysis time window can therefore be set to $[-10,50]$ ms from $t_n$, corresponding to a bounce time range of 0--60 ms. 
This procedure improves the maximal SNR of the prompt convection GW signal pixel from $\sim 3.5$ to $\sim 7.5$ (Figure \ref{fig:gcsnr}). Hence, with the coincident observation of neutrinos and GWs the detection of the GW can be claimed from the CCSN at the Galactic Center even for the case of our non-rotating progenitor. 

\begin{figure}
\begin{center}
\includegraphics[width=0.38\textwidth, bb= 50 0 550 730]{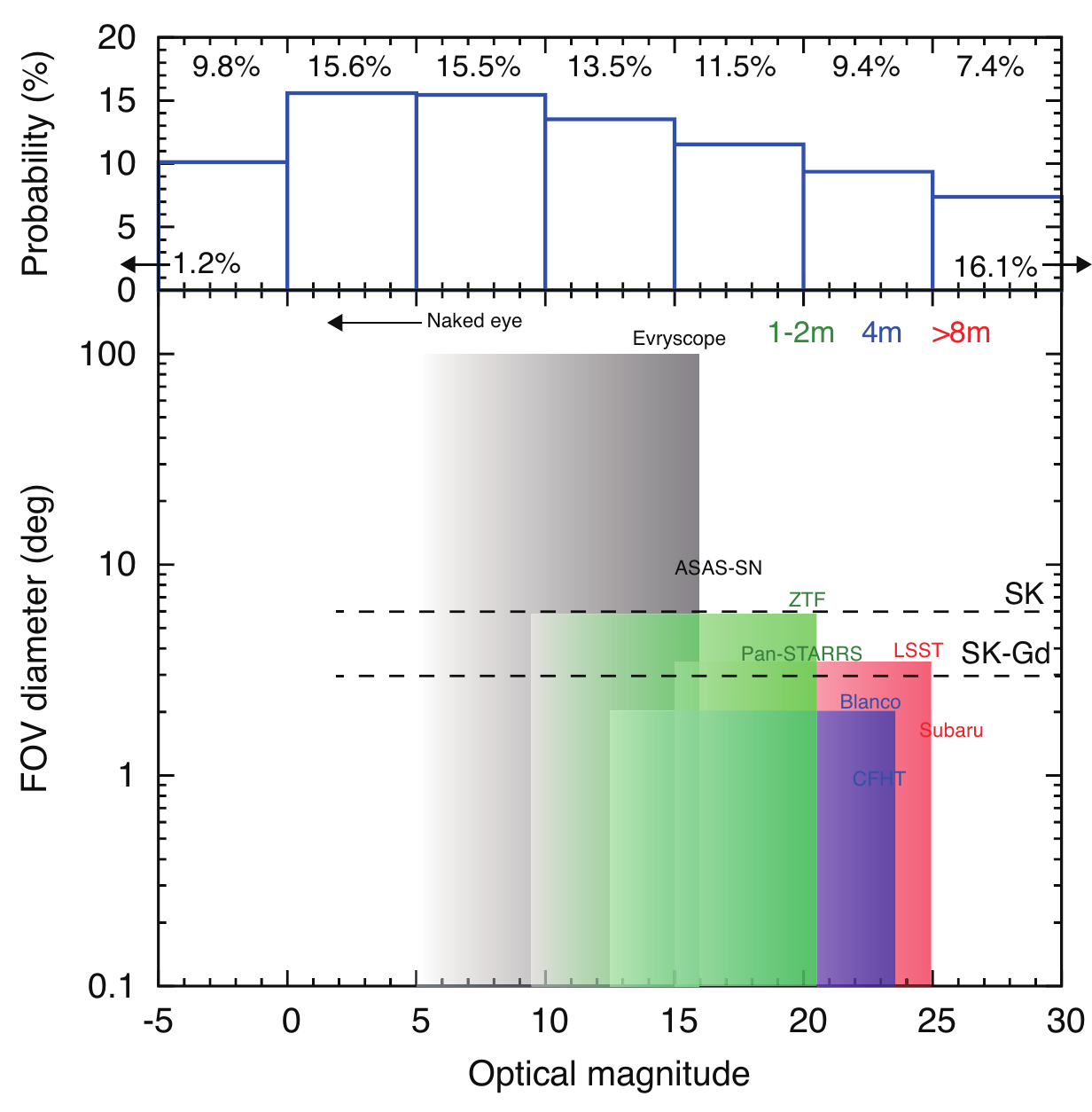} 
\caption{Detection prospects and strategies of the plateau signal of Galactic CCSNe. The top histogram shows the dust-attenuated plateau magnitudes with their respective percentage of the total CCSNe; 1.2\% and 24.5\% fall beyond the magnitude range shown. The optical magnitudes of the SBO emission are also similar to the plateau magnitudes \citep[the SBO emission is likely to be fainter by about 1 mag,][]{tominaga11}.
The bottom panel shows the typical magnitude ranges and fields of view (FOV) of various optical telescopes: ASAS-SN, Blanco, CFHT, Evryscope, LSST, Pan-STARRS, Subaru, and ZTF (shaded rectangles), as well as the naked eye (left-pointing arrow). See text for details. The error circle in CCSN pointing from the CCSN neutrino burst, for Super-K with and without Gadolinium, are represented by the horizontal dashed lines and labeled. 
}
\label{fig:opticaldetection}
\end{center}
\end{figure}

Similarly, acquiring accurate pointing information critically impacts the feasibility of rapid electromagnetic signal follow-up (Figure \ref{fig:opticaldetection}). 
As discussed above, detections in NIR wavelengths is relatively straightforward, i.e., it is almost always brighter than 5 mag. Such a bright emission can be detected with a small-aperture telescopes, whose field of views are sufficiently large in general.

In optical wavelengths, however, the expected dust attenuation makes the early detection of the EM signals challenging. The upper panel of Figure \ref{fig:opticaldetection} shows the brightness distribution of the plateau emission in optical. Approximately 25 \% of Galactic CCSNe will have apparent magnitudes $> 25$ mag, which is difficult to observe even with the largest (8 m) optical telescopes. A further 9\% will be$\sim 20$--25 mag, and require a 4--8 meter class telescope, while the dominant 40\% with $\sim 5$--20 mag will require 1--2 meter class telescopes.

The SBO emission in the optical wavelength is similar to the plateau brightness (the SBO is expected to be fainter only by about 1 mag), and thus, the upper panel of Figure \ref{fig:opticaldetection} roughly captures the brightness distribution of the SBO emission. Since the duration of the SBO emission is only about 1 hr (see Section \ref{sec:multimessengeremission} and Appendix \ref{sec:appnu}), continuous monitoring within the error circle is critical not to miss the very first SBO signal. Therefore, the positional determination by neutrino should be good enough compared with the sizes of field of views, which depends on the apertures of telescopes.

The bottom panel of Figure \ref{fig:opticaldetection} shows field of views of telescopes
as a function of typical magnitude limit. When the optical magnitude is brighter than $\sim 15$ mag, the early detection is feasible thanks to the wide field of views of small-aperture telescopes. However, fainter cases are more challenging since there is no $>$ 1m telescope with a field of view larger than $\sim 6$ degree diameter (green, blue, and red regions in Figure \ref{fig:opticaldetection}). Therefore, position determination better than 6 degree diameter is critical. To this end, the improved capability of a Gd-doped Super-K (SK-Gd), which enables the localization within 3 deg diameter for
CCSNe at the Galactic Center, would be highly impactful.

Furthermore, it must be emphasized that the optical transient may follow the neutrino burst in as short as a few minutes for a collapse of a compact Wolf-Rayet star progenitor. We conclude that pre-arranged followup programs will have an important role to play in securing the rise of the optical transient.

\section{Extremely Nearby Supernovae}
\label{sec:near}

\subsection{Neutrino}
\label{sec:nearnu}

An extremely nearby CCSN opens new probes of the core-collapse phenomenon. An example is the detection of pre-CCSN neutrinos arising from silicon burning \citep{odrzywolek04}. Neutrinos are generated by nuclear processes, including beta decay and e$^\pm$ captures, as well as thermal processes including plasmon decay and pair annihilations \citep[e.g.,][]{Misiaszek:2005ax,Patton:2015sqt}. The beta decay and pair annihilation yield \textit{O}(10) MeV neutrinos, which can be detected by Hyper-K out to several kpc \citep{odrzywolek04}, and $\approx 660$ pc with 1 kton KamLAND \citep{Asakura:2015bga}. This will provide information about the pre-collapse progenitor \citep{Kato:2015faa} and act as an advanced warning of an imminent core collapse. 

\subsection{Gravitational Waves}
\label{sec:neargw}

\begin{figure}
\begin{center}
\includegraphics[width=1.15\linewidth,height=5.0cm,bb=30 0 1020 700]{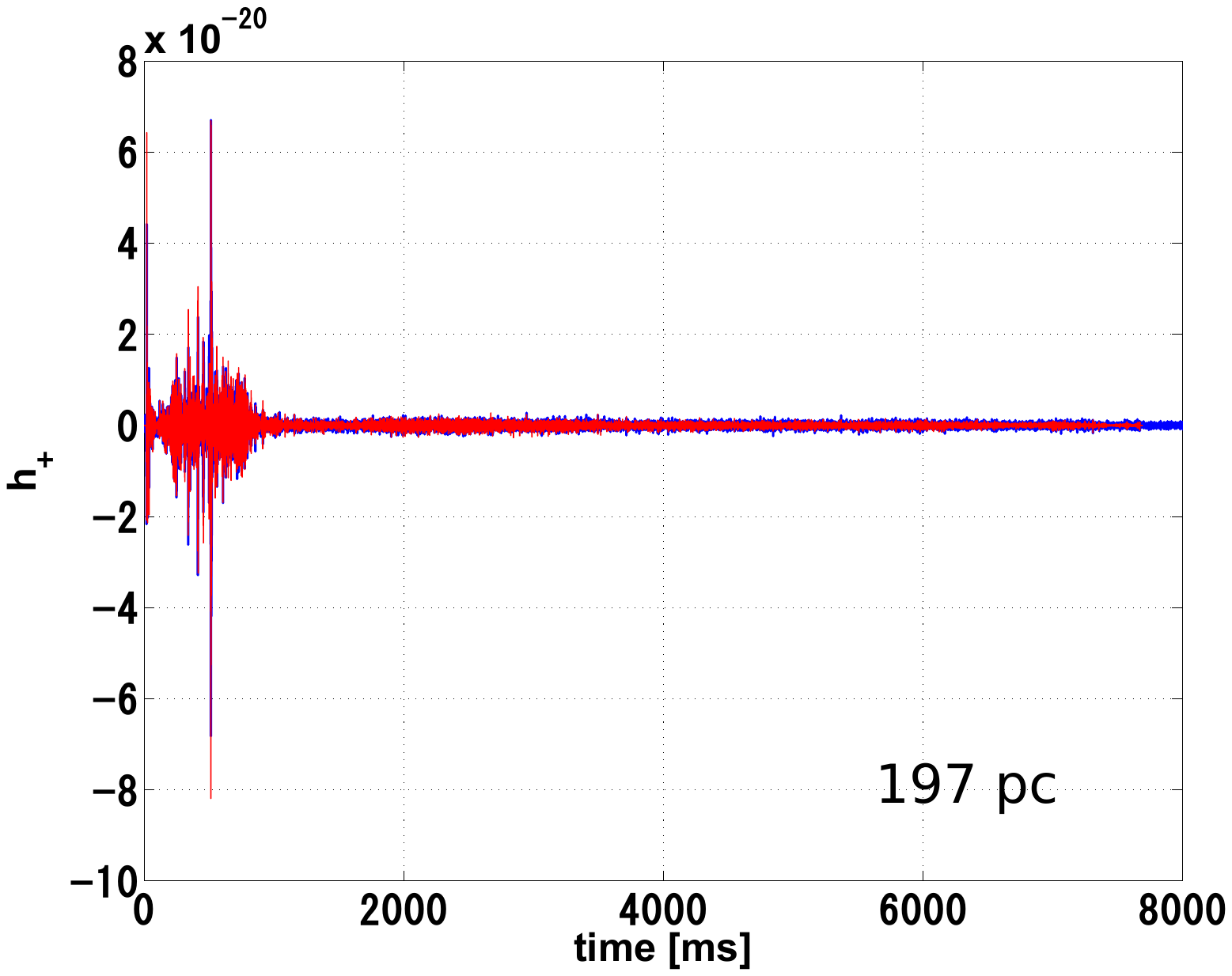} \\
\includegraphics[width=.9\linewidth,height=5.5cm,bb=50 0 1000 800]{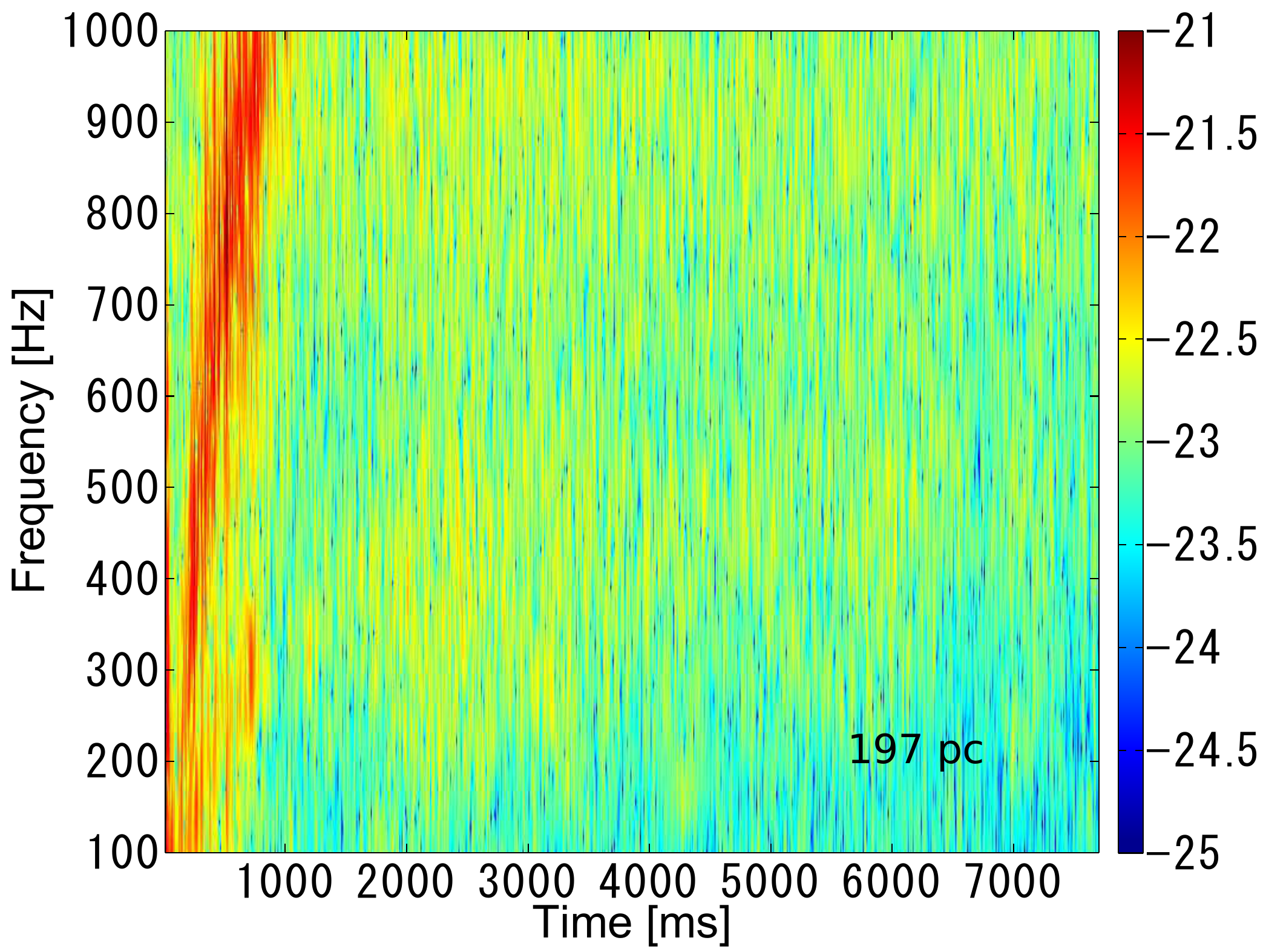} \\
\includegraphics[width=.9\linewidth,height=5.5cm,bb=50 0 1000 800]{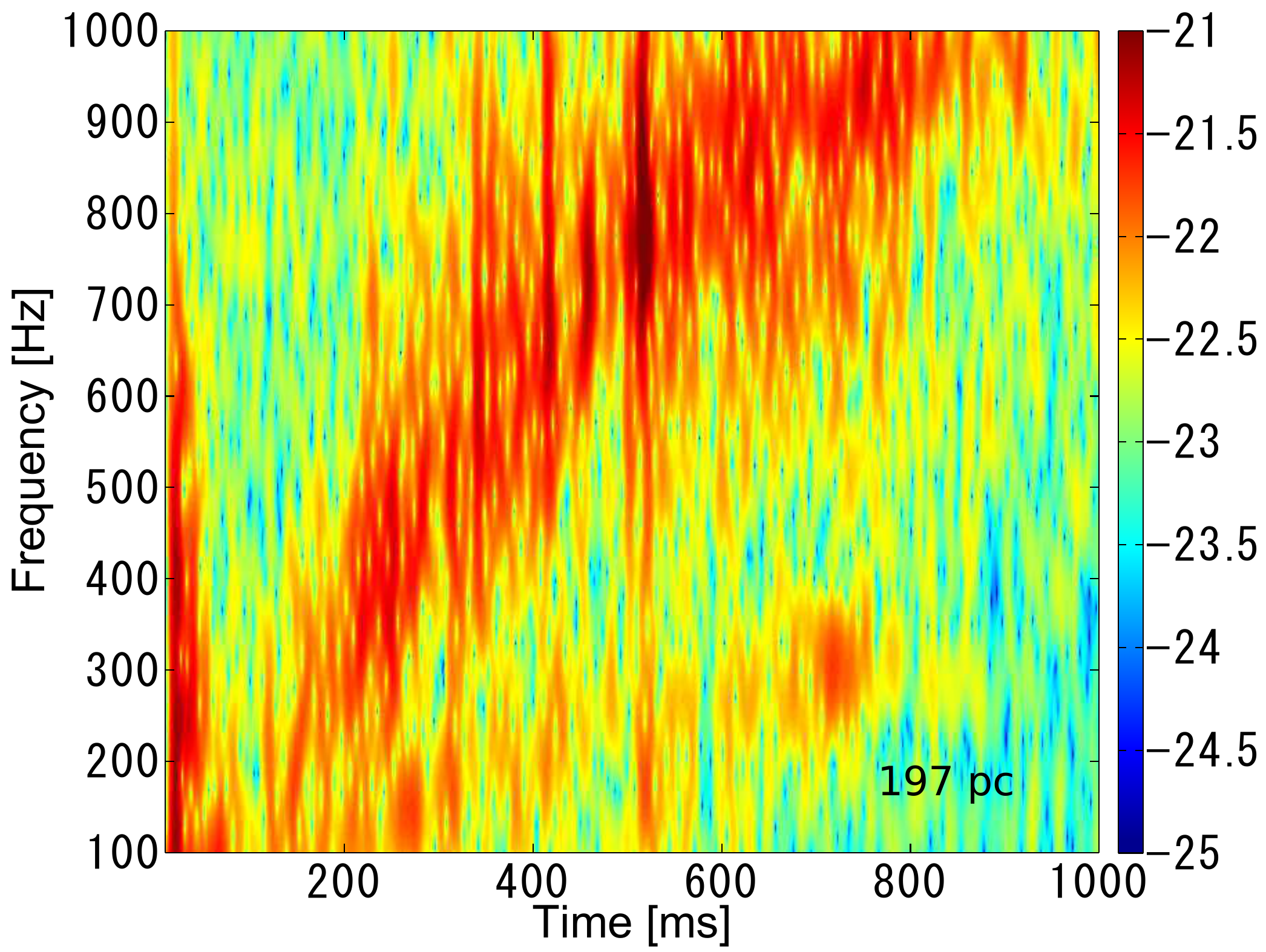}
\caption{
Reconstruction of the GW from Betelgeuse with the GW detector network H-L-V-K. 
In the top panel the red line shows the input signal while the blue line shows time series of the reconstructed signal. The reconstruction works very well and the blue line is almost hidden behind the red line.
The central and bottom panels show the spectrograms of the reconstructed signals for the first 8 seconds and 1 second, respectively.
} 
\label{fig:reconBetel1}
\end{center}
\end{figure}

The detection of a GW from an extremely nearby CCSN is easy for advanced GW detectors such as aLIGO, adVirgo, and KAGRA. As an example, we perform simulations of the reconstruction of the GW from Betelgeuse, which is known to be in a late stage of stellar evolution and going to explode as a supernova within the next million years. The right ascension and declination of Betelgeuse is 5.9 hours and 7.4 degrees, and the distance from the Earth is $197$ pc. We adopt the GW signal of our long-term simulation to be the signal from ``Betelgeuse supernova''. In our 2D model, only the $+$ mode of the polarization of the signal is considered. The generation of simulated data and the analysis is almost same as in Section \ref{sec:galgw} except that the data segment is set to $14$ s, so that the signal is in one data segment for simplicity.  

In the upper plot of the Figure~\ref{fig:reconBetel1}, 
the red and blue plots
represent the injected gravitational waveform and reconstructed time series signal, respectively. 
The reconstructed signal is shown to be matched with the injected waveform pretty well. The middle and bottom panels show the spectrogram of the reconstructed signal in a different time scale. To generate the spectrogram we set the data length of the fast Fourier transform to be $20$ ms. 
In remarkable contrast to the Galactic Center event (the right panel of Figure \ref{fig:tfgc}), such an extremely nearby supernova clearly presents time-evolving features. The main component at the early phase, spreading around $[100-700]$ Hz, corresponds to the prompt-convection signal and disappears until 100 ms after bounce. Then a monotonically increasing component appears, accompanied with a sub-signal around $[200-300]$ Hz. The increasing feature is well fitted by the Brunt-${\rm V\ddot{a}is\ddot{a}la}$ frequency at the PNS surface \citep{muellerb13}. 
The sub-signal would be originated from some hydrodynamical instabilities like SASI \citep{cerda13}.

\begin{figure*}
\begin{center}
\begin{tabular}{cc}
\includegraphics[width=0.45\textwidth,height=6cm,bb=0 0 800 600]{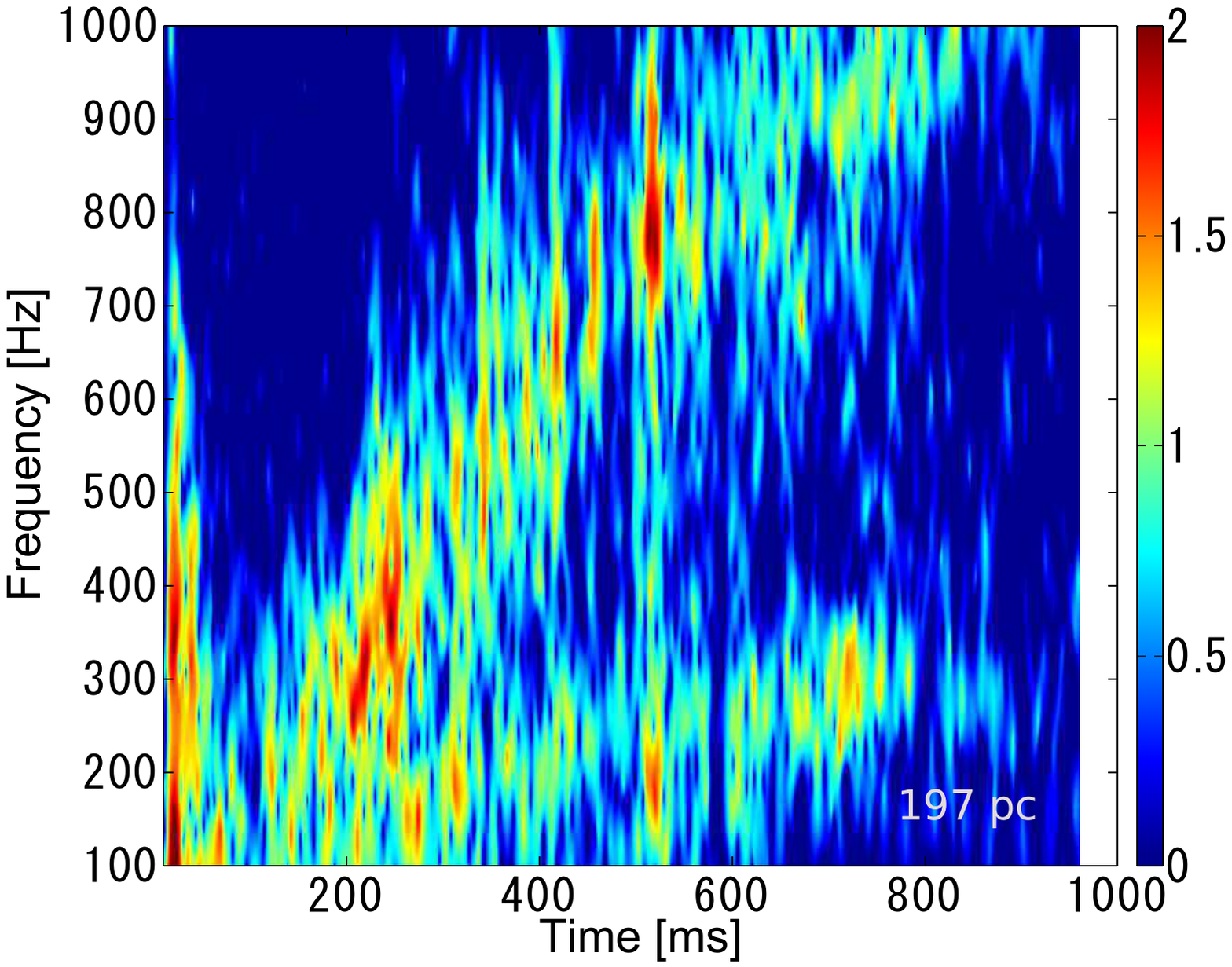} &
\includegraphics[width=0.45\textwidth,height=6cm,bb=0 0 1000 765]{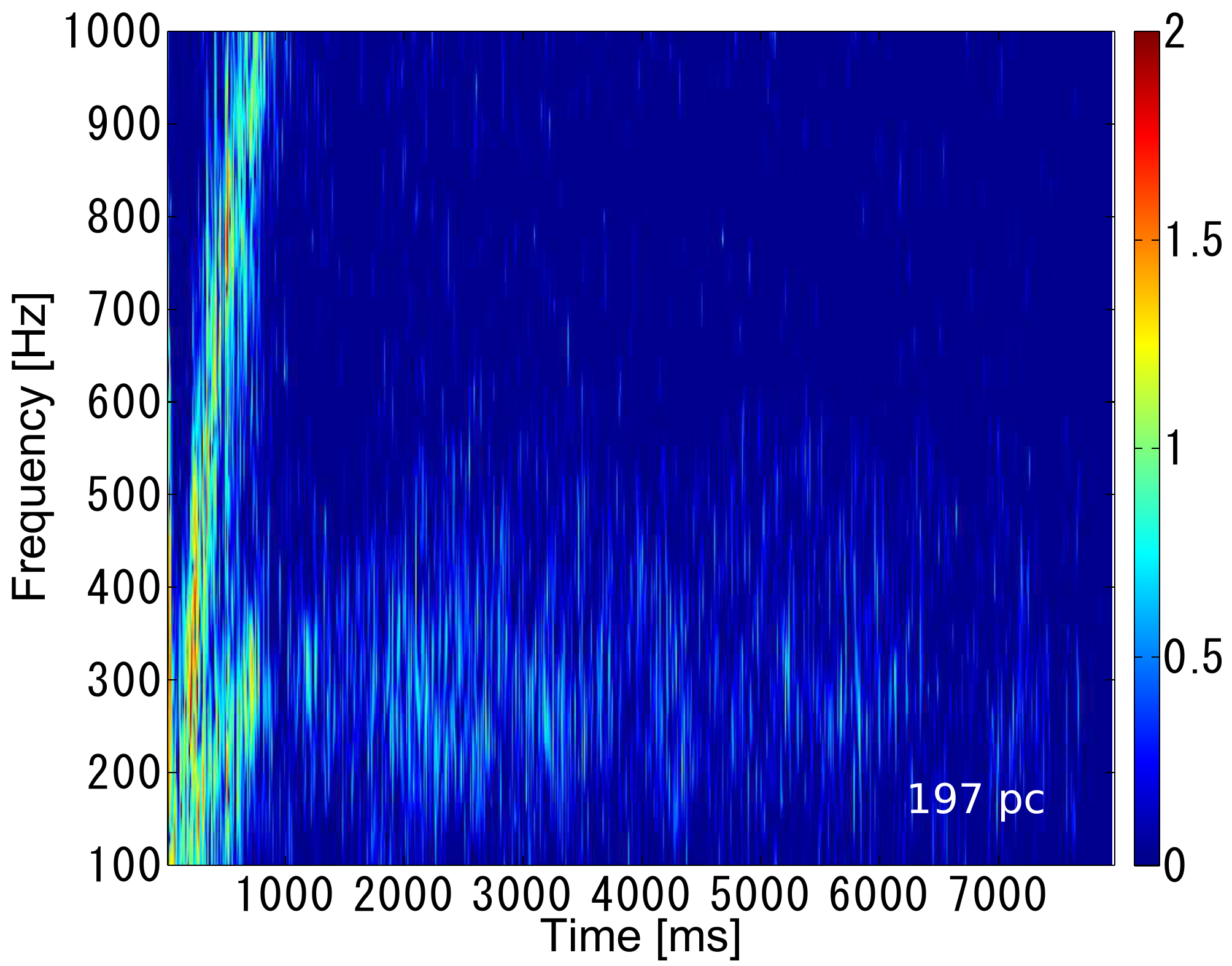}
\end{tabular}
\caption{
SNR of the GW from Betelgeuse for the first 1 second (left panel) and 8 seconds (right). The color is in logarithmic scale. 
} 
\label{fig:betelsnr}
\end{center}
\end{figure*}

We inspect the detectability of these features in the same manner as in Section \ref{sec:galgw}. The left and right panels of Figure~\ref{fig:betelsnr} present the SNR in time-frequency pixels of the first 1 s and 8 s, respectively. In the first 1 s, the signals from the prompt-convection are clearly detectable with SNR $> 100$. The post-prompt-convection signals become outstanding at $\sim 200$ ms. There are some spots with SNR 10 -- 100 up to $\sim 800$ ms. Thereafter the GW energy is distributed to  broad-band frequency regions and the energy of each time-frequency tile decreases. The reconstructed signals, however, still retain the time-frequency structure around 200--300 Hz up to $\sim 6.5$ s, of which SNR is 2--5. 

\subsection{Electromagnetic waves}
\label{sec:nearem}

It is obvious that an extremely nearby CCSN within $\lesssim 1$ kpc 
will become extremely bright both in optical and NIR wavelengths.
For 1 kpc distance (distance modulus of 10 mag),
the brightness of the plateau is expected to be about $-6$ mag.
This is as bright as the recorded brightness of the Crab nebula \citep{stephenson02}.
As is the case for SN 1054 or the Crab nebula,
such CCSN explosion would be visible even in daytime.
Thanks to the close distance, it is unlikely that the brightness
is heavily affected by the interstellar extinction.

\subsection{Multi-messenger observing strategy}
\label{sec:nearmm}

Extremely nearby CCSNe would provide
unique information on CCSNe as well as pre-CCSN stellar evolution,
and therefore, it is important not to miss any possible observables.
Thanks to the pre-CCSN neutrino signal, we have a pre-CCSN alert for 
extremely nearby events. For CCSN neutrinos and GWs,
such an alert is critical to prepare the detectors
in case the detectors are not running due to, for example,
maintenance works.

For electromagnetic wave signals, the pre-CCSN signal provides
an unique opportunity to observe massive stars just before their collapse.
Although the positional localization with pre-CCSN neutrinos is
difficult, there are not too many massive stars close to the sun,
and thus, we can keep monitoring these stars
after the detection of pre-CCSN neutrinos.
Note that the small number of massive stars also means
a low probability of CCSN rate.
For example, the CCSN rate within 660 pc 
(detection range of 1 kton KamLAND)
and 3 kpc (future Hyper-K) 
is about 0.1 and 4 \%\ of the Galactic CCSN rate, respectively.

In Table \ref{tab:RSG}, we compiled a list of nearby RSGs from the literature.
RSGs associated with OB associations are taken from
\citet{levesque05}, \citet{humphreys78}, and \citet{garmany92},
and other RSGs are supplemented from \citet{humphreys70},
\citet[Table 6]{humphreys72}, \citet{white78},
\citet[Table 20]{elias85}, and \citet[Tables 1 and 3]{jura90}.
Note that it is difficult to distinguish RSGs
and asymptotic giant branch (AGB) stars completely,
since they have an overlap in luminosity (see e.g., \citealt{levesque10}).
For the purposes of listing possible CCSN progenitor candidates,
we have chosen our selection to be rather inclusive.
The list thus includes stars with luminosity class of II (or sometimes III),
when the stars have been treated as RSG candidates in the literature,
and the list may also include intermediate-mass stars.
Lists of Wolf-Rayet stars, which are the progenitors of stripped envelope (H-poor) 
CCSNe, 
can be found in, e.g., \citet{vanderHucht01} and \citet{rosslowe15}.

Our final RSG list consists of 212 RSG candidates.
The typical distance limit is $\sim3$ kpc,
which is fortunately similar to the distance limit for
pre-CCSN neutrino detection with future neutrino detectors.
These progenitor stars and their following electromagnetic CCSN signals
are bright enough that monitoring observations soon after the pre-CCSN 
neutrino detection is feasible with 
small telescopes 
(Figure \ref{fig:opticaldetection}).

\begin{table*}
\caption{List of nearby RSG candidates}
\begin{tabular}{lllllllll} 
\hline
  Name     &  RA        & Dec        & Distance & $V$ mag  & Spec. type & Note  & Type ref$^{a}$ & Dist. reff$^{b}$ \\
           & (J2000.0)  & (J2000.0)  &  (kpc)   &          &                 &       &          &   \\
\hline
             BD+61 8 & 00:09:36.37 & $+$62:40:04.1 &  2.40  &  9.49 &     M1ep Ib + B &    KN Cas         &   1 &   2      \\ 
            BD+59 38 & 00:21:24.29 & $+$59:57:11.2 &   2.09   &  9.67 &        M2 I     &    MZ Cas         &    1 &  1       \\ 
           HD 236446 & 00:31:25.47 & $+$60:15:19.6 &  2.40   &  8.71 &     M0 Ib       &                   &  1 &  3       \\ 
              TY Cas & 00:36:59.42 & $+$63:08:01.7 &  2.40   &  11.5 ($B$)  &     M6          &                   &  1 &  3       \\ 
            V634 Cas & 00:49:33.53 & $+$64:46:59.1 &  2.51   & 10.46 &     M1 Iab      &                   &  1  &  3       \\ 
             HD 4817 & 00:51:16.38 & $+$61:48:19.8 &  1.05   &  6.18 &     K5 Ib       & HR 237            &  4  &  4      \\ 
             HD 4842 & 00:51:26.00 & $+$62:55:14.9 &  2.51   &  9.62 &     M6/7III     &    VY Cas         &  1 &  **      \\ 
           BD+62 190 & 01:03:15.35 & $+$63:05:10.8 &  2.51   &  9.95 &     M5?         &                   &  1 &  **      \\ 
           BD+62 207 & 01:08:19.93 & $+$63:35:11.2 &  2.51   &  9.82 &      M4 Iab     &    HS Cas         &   1 &   2       \\ 
           HD 236697 & 01:19:53.62 & $+$58:18:30.7 &   2.51   &  8.62 &        M1.5 I   &    V466 Cas       &    1 &  1       \\ 
           \hline
\multicolumn{9}{l}{Only the first ten rows are shown in this table. The full list is available in online material.}\\
\multicolumn{7}{l}{\textit{http://th.nao.ac.jp/MEMBER/nakamura/2016multi/}}\\
\multicolumn{9}{l}{$^a$ References for the spectral type.}\\
\multicolumn{9}{l}{$^b$ References for the distance. }\\
\multicolumn{9}{l}{References:}\\
\multicolumn{9}{l}{RSGs in OB associations: 1 \citet{levesque05}, 2 \citet{humphreys78}, 3 \citet{garmany92},}\\
\multicolumn{9}{l}{Other RSGs: 4 \citet{humphreys70}, 5 \citet{humphreys72}, 6 \citet{white78}, 7 \citet{elias85},}\\
\multicolumn{9}{l}{8 \citet{jura90}}\\
\multicolumn{9}{l}{ Additional references for the spectral type: * \citet{keenan89}, ** classification in SIMBAD}\\
\multicolumn{9}{l}{Additional references for the distance: $\dagger$ \citet{harper08},  $\ddagger$ \citet{choi08} }\\
\label{tab:RSG}
\end{tabular}
\end{table*}

\section{Extragalactic Supernovae}
\label{sec:extra}

\subsection{Neutrino}
\label{sec:extranu}

Due to its large volume, Hyper-K opens the search for neutrinos from CCSNe occurring in nearby galaxies beyond the Milky Way and Andromeda Local Group \citep{Ando:2005ka}. Since the signal is not a statistically overwhelming burst, the detectability of the signal depends crucially on backgrounds. The main backgrounds in the \textit{O}(10) MeV energy range for pure water detectors are those due to invisible (sub-\v{C}erenkov) muons decays, atmospheric neutrinos, and spallation daughter decays \citep{Bays:2011si}. 

\begin{figure}
\begin{center}
\includegraphics[width=0.3\textwidth, bb=150 0 630 600]{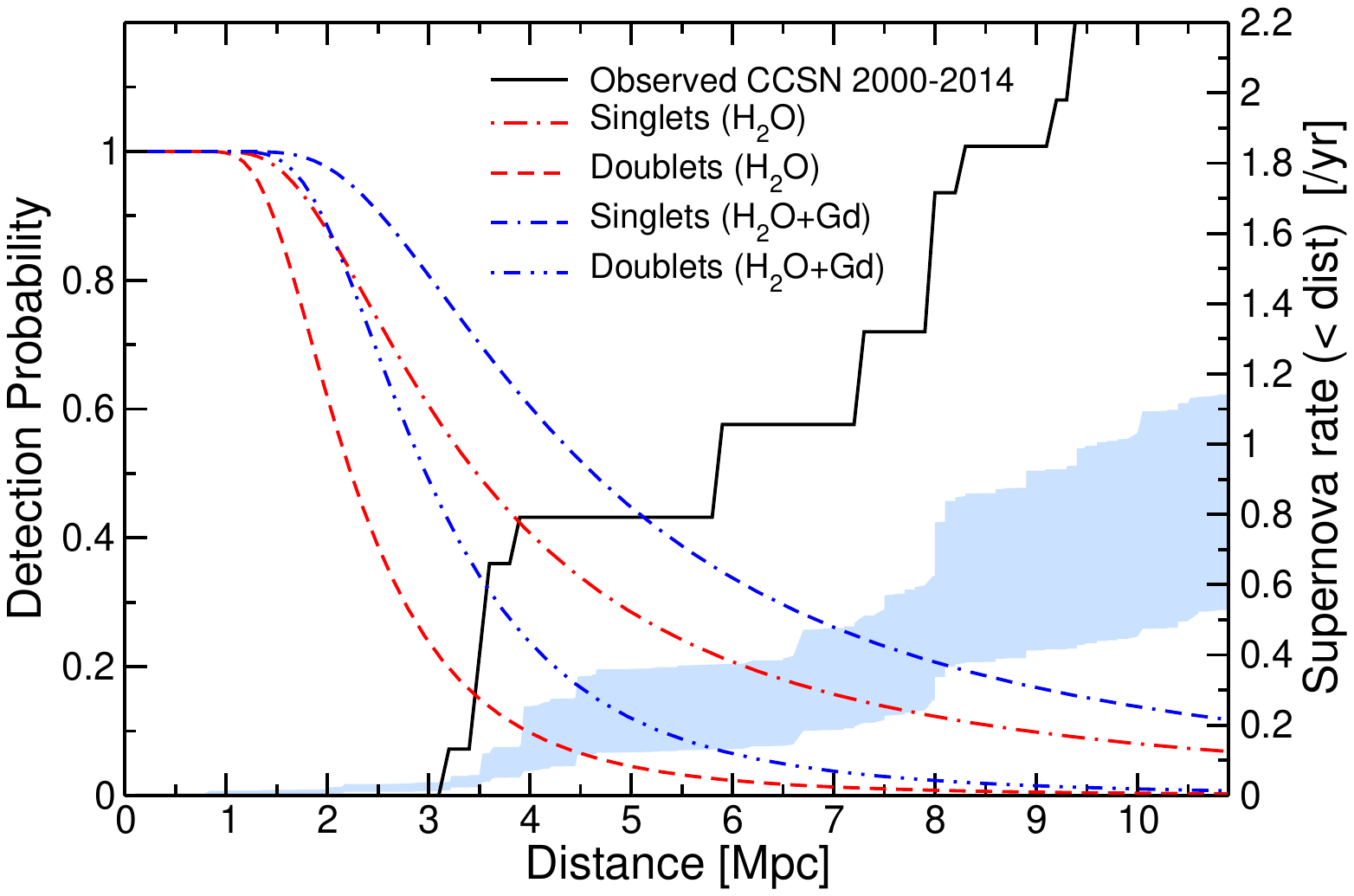} 
\caption{The probability of detecting at least a neutrino singlet (dot-dashed red) and at least a doublet (dashed red) from a CCSN of distance $D$ away with Hyper-K (left axis). Positron energy bin 18--30 MeV has been assumed. The equivalents with a Gadolinium-doped Hyper-K are shown by blue lines as labeled. On the same plot, the cumulative CCSN rate within distance $D$ is shown (right axis), estimated from observed CCSNe within the past 15 years (solid black) and the range of estimates based on observed galaxy $B$-band and star-formation rates (blue shaded band). The shaded band includes uncertainties arising from different indicators ($B$-band, H$\alpha$, and UV) as well as calibration factors (non-rotating and rotating stellar tracks). All estimates have been corrected for sky coverage incompleteness (see text).
}
\label{fig:local}
\end{center}
\end{figure}

Driven by these backgrounds, the optimal energy window to search for CCSN neutrinos over backgrounds in pure water lies in the positron energy $18$--30 MeV range. Gadolinium doped water reduces backgrounds and opens a larger energy range of approximately 12--38 MeV for signal search \citep{Ando:2005ka}. To estimate the predicted signal, we adopt a fiducial volume of 0.56 Mton. The time-integrated total inverse beta event in the 18--30 MeV energy bin for our s17.0 model is then,
\begin{equation}
N_{e^+} \simeq 8.5 \left( \frac{D}{1 \, {\rm Mpc}} \right)^{-2} \left( \frac{M_{\rm det}}{\rm 0.56 \, {\rm Mton}} \right),
\end{equation}
for the fiducial MSW mixing. Considering a full-mixing or no-mixing results in a range of 8.2--8.6 events. With the fiducial MSW mixing, the probability to detect two (or more) coincident neutrinos from a CCSN 3 Mpc away is approximately 24\%. The probability to detect one or more neutrinos is 61\%. These improve to 50\% and 81\%, respectively, for the wider energy window allowed by a Gd-doped detector. In Figure \ref{fig:local}, we show the detection probability for at least a neutrino singlet (red dot-dashed) and at least a doublet (red dashed) from a CCSN in a nearby galaxy at distance $D$. The expected event increases with Gd-doped {\it Hyper-K}, and this is represented by the larger probabilities shown by the blue lines. 

We estimate the background rate based on SK-II, which has a lower Photo-Multiplier-Tube (PMT) coverage than standard Super-K configuration and perhaps similar to the eventual Hyper-K \citep{Abe:2011ts}. After cuts, the number of remaining backgrounds was 25 events in the signal region (here defined as 18--30 MeV energy deposited and 38--50 \v{C}erenkov angle), for a fiducial volume of 22.5 kton and 794 days of exposure \citep{Bays:2011si}. This scales to a background rate of $\sim 286$ events per 0.56 Mton year. While this is a simplified estimate ignoring differences in, e.g., signal efficiency and rock overburden, Hyper-K detector capabilities are not yet finalized, and it provides a useful starting point. The accidental coincidence rate within a 10 second window where the CCSN neutrino signal will occur is thus $\sim 2 \times (286 \, {\rm yr}^{-1})^2 \times (10 \, {\rm s}) = 0.05 \, {\rm yr}^{-1} $. In other words, two (or more) coincident events within a 10 second window would provide a compelling detection of CCSN neutrinos. 

For singlet event detection, the background rate is a significant limiting factor. However, two endeavors will enhance the identification of singlet signals. The first is using the optical discovery of the CCSN as a trigger, narrowing down the time window of neutrino search. 
The background rate is $\sim 0.03$ hr$^{-1}$, and by narrowing the time window to a few hours, the background event will be a small multiple of 0.03 
(see Section \ref{sec:extraem} for more details). 
The second endeavor is background rejection due to Gadolinium doping. In pure water, the neutron left behind in IBD is captured on free protons and not registered by the detector. By doping the water with gadolinium, the gadolinium readily captures the neutron and produces a cascade of gamma-rays upon decay, which can be detected by the PMTs \citep{Beacom:2003nk}. This way, IBD and background can be separated with high efficiency. Dominant backgrounds could be reduced by a factor $\sim$5, opening up the wider energy range for neutrino searches. 

The occurrence rate of such opportunities is overlaid in Figure \ref{fig:local}, where the cumulative CCSN rate estimated directly from CCSN discoveries is shown (black solid line). We adopt the most recent 15 years, from 2000--2014 inclusive. We start with the collection of CCSNe in the past 2000--2011 studied in \cite{Horiuchi:2013bc}, and update it with recent discoveries: SN~2012A in NGC~3239, SN~2012aw in NGC~3351, SN~2014ec in NGC~3351, and SN~2013ej in NGC~628. For all, we take distances from the The 11Mpc H$\alpha$ and Ultraviolet Galaxy Survey \citep[11HUGS;][]{Kennicutt:2008ce} catalog of galaxies. There has been on average one CCSN per year within $\sim 6$ Mpc. Put another way, pure water Hyper-K will will have some $\sim 20$\% chance of detecting neutrino doublets or greater every $\sim 3$ years.

The CCSN rate can also be estimated from local galaxy catalogs. For this, we adopt the CCSN estimates based on galaxy $B$-bands and galaxy type as reported by \cite{Li:2010kd}, as well as four estimates based on galaxy star-formation rate measurements: two indicators, H$\alpha$ and UV, each with two stellar evolutionary tracks, non-rotating and rotating. For all estimates, values were first obtained for a well-surveyed patch of sky corresponding to the survey area of the Local Volume Legacy Survey \citep[LVL;][]{Dale:2009zm}, then corrected for the sky that were not surveyed. The sky-coverage correction factor is $1.98$ \citep{Dale:2009zm}. Since the scatter between different estimates is larger than the formal uncertainties of each estimate, we opt to show a band which encapsulates the estimates from our five methods. Figure \ref{fig:local} shows that, as pointed out previously \citep{Ando:2005ka,Kistler:2008us,Horiuchi:2013bc}, the directly estimated CCSN rate is larger than indirect estimates.

\subsection{Electromagnetic waves}
\label{sec:extraem}

\begin{table*}
\caption{List of local galaxies within 5 Mpc ordered by their expected CCSN rates. }
\begin{tabular}{lcccccc}
\hline
Name & RA [$^\circ$] & dec [$^\circ$] & Dist [Mpc] & log($L_{H\alpha}$) & Abs.~$B$-band &  CCSN rate [yr$^{-1}$]  \\ 
(1)	& (2)		& (3)		& (4)		& (5)		& (6)		& (7) \\
\hline
NGC5236       	&	204.253	&	 -29.866	&	  4.47	&	 41.25	&	-20.26	&	 0.024 \\
NGC253        	&	011.888	&	 -25.288	&	  3.94	&	 40.99	&	-20.00	&	 0.013 \\
NGC3034       	&	148.968	&	 +69.680	&	  3.53	&	 41.07	&	-18.84	&	 0.012 \\
NGC5128       	&	201.365	&	 -43.019	&	  3.66	&	 40.81	&	-20.47	&	 0.009 \\
NGC3031       	&	148.888	&	 +69.065	&	  3.63	&	 40.77	&	-20.15	&	 0.008 \\
Maffei2       	&	040.479	&	 +59.604	&	  3.30	&	 40.76	&	-20.22	&	 0.008 \\
UGC2847       	&	056.704	&	 +68.096	&	  3.03	&	 40.65	&	-20.58	&	 0.007 \\
NGC4945       	&	196.365	&	 -49.468	&	  3.60	&	 40.75	&	-19.26	&	 0.006 \\
NGC2403       	&	114.214	&	 +65.603	&	  3.22	&	 40.78	&	-18.78	&	 0.006 \\
NGC4449       	&	187.047	&	 +44.093	&	  4.21	&	 40.71	&	-18.17	&	 0.005\\
\hline
\multicolumn{7}{l}{Only the first ten rows are shown in this table. The full list is available in online material.}\\
\multicolumn{7}{l}{\textit{http://th.nao.ac.jp/MEMBER/nakamura/2016multi/}}\\
\multicolumn{7}{l}{The CCSN rates are derived from their $B$-band magnitudes and the LICK SNuB rates. }\\
\multicolumn{7}{l}{The galaxy catalog is based on the compilation of \cite{Karachentsev:2013ipr}.}\\
\multicolumn{7}{l}{The columns show (1) the galaxy name, (2,3) RA and declination, (4) distance, }\\
\multicolumn{7}{l}{(5) log of H$\alpha$ luminosity, (6) absolute $B$-band magnitude, (7) derived CCSN rate.} \\
\label{tab:galaxies}
\end{tabular}
\end{table*}

We do not discuss the electromagnetic wave signals from extragalactic events in detail as they are already well observed. From the multi-messenger point of view, the detection of neutrinos from extragalactic events is a significant step for CCSN studies. As discussed in Section \ref{sec:extranu}, for singlet neutrino events, we can reach several Mpc with Hyper-K. In order to firmly associate singlet neutrino detections with CCSNe, we need very early detection of electromagnetic emission, within $\lesssim 1$ days after the explosion. For exceptionally well-observed CCSN, the early light curve can be used to estimate the collapse time to within a few hours \citep[e.g.,][]{Cowen:2009ev}, 
reducing the number of single background events at neutrino detectors. 
In other words, electromagnetic surveys act as triggers and background reduction for neutrino searches.

For this purpose, the entire sky area should be densely monitored not to miss any CCSNe within $\sim 5$ Mpc. Recent surveys of transients make it possible to obtain early light curves on a more regular basis. Since the absolute magnitude of the very early phase ($< 1$ day) of CCSN is about $-14$ to $-16$ mag in optical \citep{tominaga11}, it will be as bright as $14.5 - 12.5$ mag at 5 Mpc (distance modulus of $28.5$ mag). This can be detected with small aperture telescopes with $<1$m diameter. Ongoing projects, such as 
ASAS-SN \citep[][]{shappee14} 
and Evryscope \citep{law15}, will be suitable to monitor a wide area, and to detect the very early phase of very nearby CCSNe (Figure \ref{fig:opticaldetection}). These will provide a significantly better census of CCSNe in nearby galaxies, each with very early light curves. In the following section, we discuss the observing strategy in more detail.

\subsection{Multi-messenger observing strategy}
\label{sec:extramm}

\begin{figure}
\begin{center}
\includegraphics[width=0.4\textwidth, bb = 50 50 500 570]{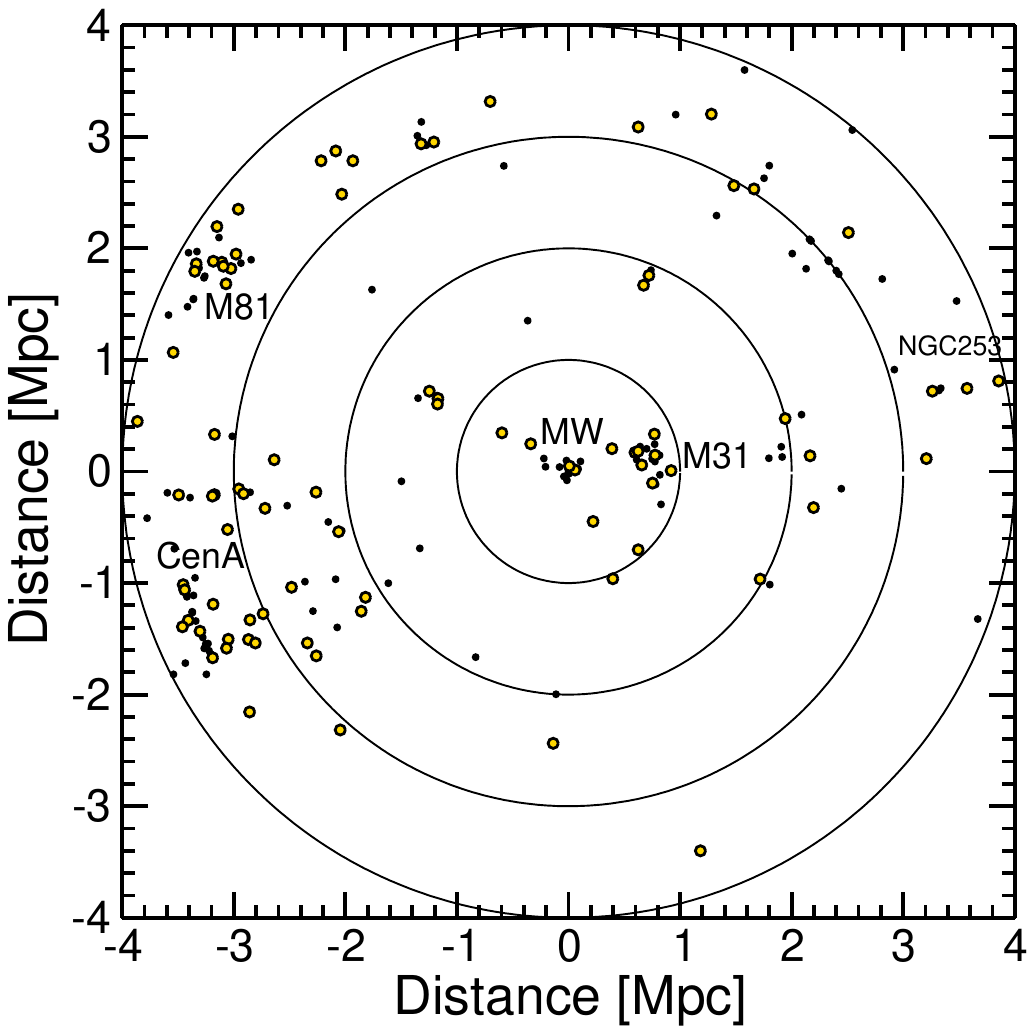} 
\caption{Sky plot showing the positions of nearby galaxies within 4 Mpc of the Milky Way based on the galaxy catalog of \citet{Karachentsev:2013ipr}. Clusters of galaxies in the M81 and Cen A clusters are labeled. M31, the nearest neighboring spiral galaxy, and NGC 253, a prominent star-forming galaxy, are labeled. }
\label{fig:skyplot}
\end{center}
\end{figure}

With only several neutrino events, the neutrino burst from an extragalactic CCSN will reveal IF to look, but it is challenging to determine WHERE to look. One method to narrow down the survey area is to use a precompiled list of galaxies. For this purpose, the nearby galaxy catalog of \cite{Karachentsev:2013ipr} is the most complete. Figure \ref{fig:skyplot} shows the distribution of nearby galaxies in right ascension (the declination has been compressed). We label a few of the prominent clusters of galaxies. These structures explain the origin of sharp rise in the CCSN rate between 3 and 4 Mpc in Figure \ref{fig:local}. Also labeled is NGC 253, a prominent nearby star-forming galaxy. 

The CCSN rates of the galaxies will serve as a probabilistic guide for which galaxy hosted the CCSN, which can further strategically narrow the area to survey. To estimate the CCSN rate of each galaxy, we adopt a subset of the Karachentsev catalog for which the star formation rate can be observationally estimated. About half of the nearby galaxies have multi-band observations from Spitzer-IR, H$\alpha$, to GALEX-UV, providing diagnostics of the star formation rate. For example, the Karachentsev catalog contains 236 galaxies within 5 Mpc, compared to the H$\alpha$ catalog's 131 \citep{Kennicutt:2008ce}, the UV catalog's 154 \citep{GildePaz:2006bw}, and the IR catalog's 112 \citep{Dale:2009zm}. Fortunately, the incompleteness is most pronounced for the least massive galaxies, and does not dramatically impact the completeness of the most important galaxies. 

We use the H$\alpha$ luminosity to estimate the CCSN rate. The H$\alpha$ flux is first corrected for [NII] line contamination, underlying stellar absorption, and Galactic foreground extinction following \cite{Kennicutt:2008ce}. Table \ref{tab:galaxies} shows the derived H$\alpha$ luminosities. Internal attenuation is corrected for using the empirical scaling correction with the host galaxy $B$-band magnitude derived in \cite{Lee:2009by},
\begin{equation}
A_{H\alpha}=\left\{ \begin{array}{lll}
0.10 & M_B > -14.5\\
1.971+0.323M_B+0.0134M_B^2 & M_B \leq-14.5 .
\end{array}
\right.
\end{equation}
We adopt a H$\alpha$ to star formation calibration of $7.9 \times 10^{-42} \, {\rm M_\odot yr^{-1} / (erg \, s^{-1})^{-1}}$ \citep[e.g.,][]{Kennicutt:1998zb}, which assumes a Salpeter initial mass function and mass range $0.1$--$100 M_\odot$. This factor can vary by $\sim 20$\% by stellar evolution codes and stellar rotation \citep{Horiuchi:2013bc}. Assuming that all stars with mass range $8$--$100 M_\odot$ undergo core collapse, the star formation to CCSN rate conversion is $0.0074 /M_\odot$ for the Salpeter initial mass function. Changing the lower mass limit to $9.5 M_\odot$ as estimated by progenitor studies \citep{Smartt:2015sfa} reduces this by $\sim 20$\%. The conversion is not strongly sensitive to the upper mass cut, e.g., adopting $40 M_\odot$ instead of $100 M_\odot$ only reduces the conversion by $\sim 8$\%. The resulting CCSN rates are summarized in the final column of Table \ref{tab:galaxies}, and a full list is available in online materials. 
We find that the 10 (20) nearby galaxies with the largest expected CCSN rate within the local 5 Mpc contain more than 60\% (87\%) of the total CCSN rate in the entire 5 Mpc volume, highlighting the strategic importance of the nearby large galaxies. 

\section{Summary and Discussion}
\label{sec:summary}

In this paper, we consider multi-message signals of CCSNe (in neutrinos, GWs, and electromagnetic waves) occurring at three distance regimes (the Galactic Center, a very nearby location, and extragalactic distances). Our signal predictions are self-consistently calculated based on a long-term simulation of an axisymmetric neutrino-driven core-collapse explosion \citep{nakamura15} initiated from a non-rotating solar-metallicity progenitor with ZAMS mass of $17 \, \Msun$. We consider both current and upcoming multi-messenger detectors and investigate the physics reach of a future CCSN event. The detectability of multi-messenger signals are summarized in Table \ref{tbl:summary}, and our main findings are summarized below. 

For a CCSN occurring at the Galactic Center 8.5 kpc away, the neutrino burst will be an effective trigger for subsequent multi-messenger observations. The neutrino signal determines the time of core bounce to within several milliseconds. This high accuracy estimation of the bounce time allows the time window of GW analysis to be reduced, which greatly reduces the noise and makes it possible to obtain a signal-to-noise ratio larger than that which would be possible without the neutrino trigger (Figure \ref{fig:gcsnr}). Secondly, the neutrino signal provides pointing information. A pure water or gadolinium-doped Super-K will reveal the CCSN direction within an error circle of $\sim 6^\circ$ or $\sim 3 ^\circ$, respectively. The latter is particularly important as it corresponds to the field of view of large ($>1$ m) modern optical telescopes (Figure \ref{fig:opticaldetection}) which are necessary for observing the SBO emissions of CCSNe occurring at the Galactic Center and other side of the Milky Way (Figure \ref{fig:appmag}). In other words, the pointing accuracy will considerably facilitate optical followup.

An extremely nearby CCSN, while rare, would provide unique information of the CCSNe as well as pre-CCSN stellar evolution. The pre-CCSN neutrino signals enable us to 
diagnose the core structure of the CCSN progenitor and to prepare detectors for observing upcoming signals. 
For such an extremely nearby event, 
the high-precision waveform reconstruction can be done
(Figure \ref{fig:reconBetel1}, top panel) 
using the coherent network analysis of aLIGO, adVirgo and KAGRA. 
Ceaseless monitoring of nearby massive stars is essential not to 
miss the possible opportunities of such a rare event. To aid early followup, we compiled a list of nearby RSG candidates (Table \ref{tab:RSG}).

At the other extreme, next-generation neutrino detectors will have sensitivity to the neutrino burst from CCSN in nearby galaxies within a few Mpc. To assist 
early followup, we compiled a list of nearby galaxies (Table \ref{tab:galaxies}) including their CCSN rates estimated from the H$\alpha$-derived star formation rates. Within the horizon of next-generation detectors, the top 10 galaxies host more than 60\% of the total CCSN rate, so the number of galaxies which need followup is rather limited. 
For the GW detection, the horizon does not go beyond $\sim 100$ kpc for our 2D (non-rotating) model even using ET (Table \ref{tbl:summary}), however, it may extend to $\sim$ Mpc distance scale if non-axisymmetric
instabilities with more efficient GW emission would be captured in 3D models \citep{Kuroda14}.

Our CCSN model is based on a 2D simulation using an approximation for neutrino transport. 
These numerical aspects, as well as multi-messenger signal predictions, ultimately need to be investigated by using self-consistent three-dimensional (3D) models with long-term evolution of at least $\sim 10$ s postbounce. Considering the rapid increase in supercomputing power \citep{Kotake12_ptep}, we speculate that we could have access to such systematic 3D models in the not too distant future. In the near future, detectors such as KAGRA and LSST will go online, and over the next decades, advanced detectors such as DUNE, ET, and Hyper-K. Collectively, these detectors promise to deliver rich data from a future Galactic CCSN, which will be indispensable for unraveling the nature of the CCSN mechanism. Much theoretical investigations are necessary to prepare for this golden event, and our long-term 2D simulation with self-consistent multi-messenger signal predictions is the first step in a string of necessary studies. 
 
\section*{Acknowledgements}

We thank John Beacom and Shoichi Yamada for valuable discussions and helpful comments. This study was supported in part by the Grants-in-Aid for the Scientific Research of Japan Society for the Promotion of Science (JSPS, Nos. 24244036, 24740117,  26707013, 26870823, and 15H02075) and the Ministry of Education, Science and Culture of Japan (MEXT, Nos. 24103005, 24103006, 25103515, 26000005, 26104001,15H00788, 15H00789, 15H01039).




\bibliographystyle{mn}
\bibliography{reference}






\appendix

\section{Multi-messenger signals}
\label{sec:appsignal}

In this Appendix, we describe signals and detectors supposed in this paper. 

\subsection{Neutrinos}
\label{sec:appnu}

Neutrino signals are a direct outcome of our CCSN simulation which takes account neutrino transport in a self-consistent way. After emission from the collapsed core, neutrinos undergo flavor conversions during propagation to the Earth. In general, neutrino oscillation varies during the core-collapse evolution, and also depends on the neutrino mass hierarchy. Mikheyev-Smirnov-Wolfenstein (MSW) matter effects in the normal mass hierarchy imply that the observed $\nu_e$ and $\bar{\nu}_e$ fluxes, $F_{\nu_e}$ and $F_{\bar{\nu}_e}$, are \citep{Dighe:1999bi},
\begin{eqnarray}
F_{\nu_e}& \simeq &F^0_{\nu_x}  \label{eq:NHmixing1},  \\
F_{\bar{\nu}_e}& \simeq& \cos^2 \theta_{12} F^0_{\bar{\nu}_e} + \sin^2 \theta_{12} F^0_{{\nu}_x},  \label{eq:NHmixing}
\end{eqnarray}
where fluxes denoted by the superscript $0$ represent those emitted from the neutrinospheres and $\nu_x$ represents heavy lepton neutrinos. Here, $\theta_{12}$ is the solar mixing angle and $\sin^2 \theta_{12} \simeq 1/3$. For inverted mass hierarchy the same quantities become,
\begin{eqnarray} 
F_{\nu_e}& \simeq &\sin^2 \theta_{12} F^0_{\nu_e} + \cos^2 \theta_{12} F^0_{\nu_x} \label{eq:IHmixing1}, \\
F_{\bar{\nu}_e} &\simeq &F^0_{\bar{\nu}_x}.\label{eq:IHmixing}
\end{eqnarray}
Therefore, depending on the mass hierarchy, the observed $\bar{\nu}_e$ flux is composed of either a mix of $\bar{\nu}_e$ and $\nu_x$ at the neutrinospheres, or entirely of $\nu_x$ at the neutrinosphere. We neglect for simplicity Earth matter crossing effects. 

For much of the post-bounce evolution, one finds a flux ordering $F^0_{\nu_e} > F^0_{\bar{\nu}_e} > F^0_{\nu_x}$. This ordering allows for additional flavor mixing induced by the coherent neutrino-neutrino forward scattering potential \citep[for a review, see, e.g.,][]{Duan:2010bg,Mirizzi:2015eza}. In the inverted mass hierarchy, the observed $F_{{\nu}_e}$ and $F_{\bar{\nu}_e}$ are given by Eqs.~(\ref{eq:NHmixing1}--\ref{eq:NHmixing}) instead of Eqs.~(\ref{eq:IHmixing1}--\ref{eq:IHmixing}), but this effect is energy-dependent, giving rise to 
so-called spectral splits in the neutrino energy spectrum. 
Also, additional effects such as the ``halo effect'' become important \citep{Cherry:2012zw} and exact predictions of collective flavor transformation in the post-bounce environment, and when it will apply, and at what energies, are still under debate. 
Therefore, the observed $\bar{\nu}_e$ will be some mixture of $\bar{\nu}_e$ and $\nu_x$ at source, whose mixture will be energy- and time-dependent. We therefore adopt Eqs.~(\ref{eq:NHmixing1})--(\ref{eq:IHmixing}) as two representative cases, and also consider a scenario where $F_{{\nu}_e}  = F^0_{{\nu}_e} $ and $F_{\bar{\nu}_e}  = F^0_{\bar{\nu}_e} $ (which we will call the ``no-mixing'' scenario) to cover all extremes, as was done in Ref.~\citep{Tamborra:2014hga}.


The dominant sensitivity of Super-K and its successor Hyper-K, as well as JUNO, is to $\bar{\nu}_e$ through the inverse beta reaction $\bar{\nu}_e + p \to e^+ n$. In Super-K and Hyper-K, the targets are free protons in water; for JUNO the targets are free protons in linear alkyl benzene ($C_6H_5C_{12}H_{25}$). 
The cross section and kinematics are well-known: $\sigma(E_\nu) \simeq 0.0952 \times 10^{-42} (E_\nu-1.3)^2(1-7E_\nu/m_p) \, {\rm cm^2} $, where $E_\nu$ is the neutrino energy and $m_p$ is the proton rest mass; the threshold for the reaction is $E_\nu > 1.8$ MeV;  the kinetic energy of the positron $T_e \simeq E_\nu -1.8 $ MeV faithfully represents the neutrino energy; and the positrons are emitted almost isotropically \citep{Strumia:2003zx,Vogel:1999zy}.

Elastic scattering $\nu + e^- \to \nu + e^-$ on electrons provides a strongly complementary channel to inverse beta decay. It is sensitive to all neutrino flavors, has poor neutrino energy resolution, but good angular resolution of the incoming neutrino. The recoil kinetic energy of the forward-scattered electron varies between 0 and $2E_\nu^2/(m_e+2E_\nu)$, and makes an angle ${\rm cos} \, \alpha = \sqrt{T_e/(T_e+2m_e)}(E_\nu+m_e)/E_\nu$. The cross section is also well known \citep{Vogel:1989iv}.

We neglect interactions with oxygen, e.g., the charge-current $\nu_e + {\rm ^{16} O} \to e^- + {\rm ^{16}F^*}$ and $\bar{\nu}_e + {\rm ^{16}O} \to e^+ + {\rm ^{16}N^*}$, as well as all-flavor neutral-current channels \citep{Haxton:1987kc}. This is because they have high threshold energies and their yields are smaller compared to that from the inverse beta channel.

\subsection{Gravitational wave}
\label{sec:appgw}

We extract the GW signal from our simulations 
using the standard quadrupole formula (e.g.,
\citealt{Misner:1974qy}). For numerical convenience,
we employ the so-called first-moment-of-momentum-divergence 
(FMD) formalism proposed by \cite{finn90} (see also 
modifications in \citet{Murphy:2009dx,muellerb13}).
The axisymmetric (transverse-traceless) GW strain $h_+$ is then given by
\begin{equation}
h_+ = \frac{3}{2}\frac{G}{Dc^4} \sin^2 \alpha \frac{d^2}{dt^2}\,\, \Ibar_{zz}\,\,,
\label{fmd}
\end{equation}
where $G$ is the gravitational constant, $c$ is the speed of light, $D$ is the 
 distance to the source, and 
the reduced mass-quadrupole tensor
$\Ibar_{jk}$ is given by
\begin{equation}
\Ibar_{jk} = \int \rho 
\bigg(x_j x_k - \frac{1}{3}\delta_{jk} x_i x^i\bigg) d^3 x,
\label{Ibar}
\end{equation}
where $\rho$ is the density, $x_j$ is the spatial coordinate, and
 $\alpha$ represents the angle between the symmetry axis and the line of
sight of the observer, which we take to be $\pi/2$ to maximize the GW amplitude. 
In axisymmetric simulations, the non-vanishing component
of the second derivative in Equation (\ref{fmd}) reads
\begin{eqnarray}
\label{second}
\frac{d^2}{dt^2}\, \Ibar_{zz} &=& \frac{8\pi}{3} \int d\cos\theta \int
dr\, r^3  \\
&\times& \bigg[ P_2(\cos\theta)\, \frac{d \rho v_r}{dt} + \frac{1}{2}
\frac{\partial}{\partial \theta} P_2(\cos\theta)\, \frac{d \rho
v_\theta }{dt}\bigg]\,\,, \nonumber
\end{eqnarray}
where $P_2(\cos\theta)$ is the second Legendre polynomial in
$\cos\theta$ and $v_r$ and $v_\theta$ are the fluid velocities in the
radial and lateral direction, respectively \citep{Murphy:2009dx}.
The time derivative in the right hand side of Equation (\ref{second})
is numerically evaluated in the simulation.

The total emitted energy in GWs is estimated by
\begin{equation}
E_\mathrm{GW} = \frac{3}{10} \frac{G}{c^5} \int
\bigg(\frac{d^3}{dt^3}\, \Ibar_{zz}\bigg)^2\,dt\,,
\label{eq:egw}
\end{equation}
and the GW spectral energy density via 
\begin{equation}
\frac{d E_\mathrm{GW}}{df} = \frac{3}{5}\frac{G}{c^5} (2\pi f)^2 |\tilde{A}|^2\,,
\label{eq:spect}
\end{equation}
where $\tilde{A}(f)$ denotes the Fourier transform of $A \equiv
\frac{d^2}{dt^2}\, \Ibar_{zz}$. Then the
 characteristic GW strain ($h_{\rm char}$)
 \citep{Flanagan:1997sx} can be defined as
\begin{equation}
\label{eq:hchar}
h_\mathrm{char} = \sqrt{\frac{2}{\pi^2} \frac{G}{c^3} 
\frac{1}{D^2} \frac{dE_\mathrm{GW}}{df}}\,\,.
\end{equation}


We use the coherent network analysis for GW detection. 
The essence is as follows. 
Suppose a GW interacts with $m$ interferometric detectors. 
The response of the $I$\mbox{-}th detector is written as 
$\xi_I (t) = \Sigma_A F_I^A h_A(t)$ 
with $h_A (t)$ representing the GW signal; 
here the subscripts run as $I=1,\cdots, m$ and $A=+,\, \times$, 
the latter of which represents the plus~($+$) 
and cross~($\times$) polarization components of the GW, respectively; 
$F_{I+}(\hat{\Omega}_I)$ and $F_{I\times}(\hat{\Omega}_I)$ are 
the antenna patterns of the $I$\mbox{-}th detector 
for a given sky direction $\hat{\Omega}_I:=(\phi_I,\theta_I)$. 
Data $x_i(t)$ from the $I$\mbox{-}th detector are expressed as
\begin{equation}
x_I(t)=\xi_I(t+\tau_I)+\eta_I(t), \quad \tau_I=\tau(\phi_I,\theta_I),
\label{eq:data}
\end{equation}
where $\tau_I$ is a relative time delay with respect to the center of the Earth, 
which depends on the source location $(\phi_I,\theta_I)$, 
and $\eta_I(t)$ is the detector noise. 
Since noise from an interferometric GW detector 
has frequency dependence, 
we usually apply a whitening filter to the data. 
  
Shifting the data time $t+\tau_I \mapsto t$, 
we rewrite equation (\ref{eq:data}) as 
$x_I(t_I)=\xi_I(t)+\eta_I(t_I), t_I=t-\tau_I$. 
All detectors being taken into account, we obtain
\begin{eqnarray}
\boldsymbol{x} = \boldsymbol{F}\boldsymbol{h} + \boldsymbol{\eta},
\label{eq:matrixformula}
\end{eqnarray}
where 
$\boldsymbol{x}=(\boldsymbol{x_1},\cdots,\boldsymbol{x_m})^T \in W[m \times N]$, 
$\boldsymbol{F}=(\boldsymbol{F_+},\boldsymbol{F_\times}) \in W[m \times 2]$, 
$\boldsymbol{h}=(\boldsymbol{h_+},\boldsymbol{h_\times})^T \in W[2 \times m]$, 
$\boldsymbol{\eta} = (\boldsymbol{\eta_1},\cdots,\boldsymbol{\eta_m})^T \in W[m \times N]$. 
Fourier-transforming these quantities as 
$\boldsymbol{x_I}=(\tilde{x_I}(f_0),\cdots,\tilde{x_I}(f_{N-1})) \in W[1 \times N]$, 
$\boldsymbol{h}_A=(\tilde{h}_A(f_0),\cdots,\tilde{h}_A(f_{N-1})) \in W[1 \times N]$, 
$\boldsymbol{\eta}_I=(\hat{\eta}_I(f_0),\cdots,\hat{\eta}_I(f_{N-1})) \in W[1 \times N]$, 
we define the whitened counterparts as 
$\hat{x}_I(f_j):=\tilde{x}_I(f_j)/\sqrt{S_\mathrm{n}^I(f_j)}$, 
$\hat{F}_I^A(\hat{\Omega},f_j):=F_I^A(\hat{\Omega})/\sqrt{S_\mathrm{n}^I(f_j)}$, 
and 
$\hat{\eta}_I(f_j):=\tilde{\eta}_I(f_j)/\sqrt{S_\mathrm{n}^I(f_j)}$, 
$j=0,\cdots,N-1$ 
with the power spectrum density $S_\mathrm{n}^I(f_j)$ 
of the $I$-th detector noise at the frequency $f_j$. 

Since $\boldsymbol{F}$ is not a square matrix of full rank in general, 
the Moore-Penrose pseudo-inverse matrix 
$\boldsymbol{M}:=\boldsymbol{F}^T\boldsymbol{F}$ 
is employed to obtain 
$\boldsymbol{M}\boldsymbol{h}=\boldsymbol{F}^T\boldsymbol{x}$. 
If the detectors are not all co-aligned, 
$\boldsymbol{M}$ is an invertible $m\times m$ matrix, 
and we get 
$\boldsymbol{h}=\boldsymbol{F}^{\dagger}\boldsymbol{x}$, 
where $\boldsymbol{F}^\dagger:=\boldsymbol{M}^{-1}\boldsymbol{F}^T$.
  
The maximum likelihood method is then applied 
to reconstruct the signal $\boldsymbol{h}$. 
The likelihood function is given as
$p(\boldsymbol{x_I}|\boldsymbol{h})\propto \exp\left(-\frac{1}{2}\sum_j|\boldsymbol{x_I}(j\Delta_f)-\boldsymbol{F_I}\boldsymbol{h}(j\Delta_f)|^2\right)$ 
where $\Delta_f$ is the frequency resolution.
Taking the logarithm of the likelihood ratio for convenience, 
we define:
\begin{eqnarray}
	&&L[\boldsymbol{h}]:=\log\Bigl(\frac{p(\boldsymbol{x_I}|\boldsymbol{h})}{p(\boldsymbol{x_i}|\boldsymbol{0})}\Bigr)\label{eqn:gaussLR}\\
	    &=&
	    \sum_j\frac{1}{2}\left(\left|\boldsymbol{x_I}(j\Delta_f)\right|^2 \,-\, \left|\boldsymbol{x_I}(j\Delta_f)-\boldsymbol{F_I}\boldsymbol{h}(j\Delta_f)\right|^2\right)\nonumber
\label{eq:likelihood}
\end{eqnarray}	
The maximum of $L[\boldsymbol{h}]$ is then found 
by varying $\theta$ and $\phi$ over their parameter spaces. 
The reconstruction of the polarization waveforms is finally achieved as 
$\boldsymbol{h}_A=\boldsymbol{H}_A\cdot\boldsymbol{x}$, 
where
$\boldsymbol{H}_A = \frac{1}{\mathrm{det}(\boldsymbol{M})} \sum_{A'} \boldsymbol{C}_{AA'} \cdot \boldsymbol{F}_{A'}$, $\boldsymbol{C}_{AA'}=[\mathrm{adj}(\boldsymbol{M})]_{AA'}$, $\mathrm{adj}(\boldsymbol{M})$ 
is the adjoint matrix 
and $\mathrm{det}(\boldsymbol{M})$ is the determinant of $\boldsymbol{M}$.

\subsection{Electromagnetic waves}
\label{sec:appem}

The first electromagnetic signal from the CCSN is 
the emission from SBO, which occurs 
when the supernova shock reaches the surface of the progenitor. 
Therefore, for the case of a RSG progenitor as discussed in this paper, 
the first electromagnetic signal will rise about 1 day 
after the neutrino and GW emissions. 
This delay enables neutrino and GW signals 
to be used as precursors for the SBO signal. 
However, it should be emphasized that this delay can be as 
short as a few minutes for the case of Wolf-Rayet star progenitors
that have significantly smaller stellar radii.

Given the large difference in time scales involved, it is difficult for 
current simulations of core collapse to directly reproduce the SBO 
signal. Therefore, we employ analytic expressions
\citep{matzner99} for the SBO luminosity and duration,
\begin{eqnarray}
L_{\rm SBO} &\sim& 2.2 \times 10^{45} \, {\rm ergs \ s}^{-1}
         \left( \frac{\kappa}{0.34\ {\rm cm^2 \ g}} \right)^{-0.29} \nonumber \\
         & \times & \left( \frac{M_{\rm ej}}{10 \, M_\odot} \right)^{-0.65}
         \left( \frac{E_{\rm exp}}{10^{51} \, {\rm erg}} \right)^{1.35}
         \left( \frac{R_0}{500 \, R_\odot} \right)^{-0.42}
\end{eqnarray}
and
\begin{eqnarray}
t_{\rm SBO} & \sim& 790 \ {\rm s}
         \left( \frac{\kappa}{0.34\ {\rm cm^2 \ g}} \right)^{-0.58} \nonumber \\
         & \times & \left( \frac{M_{\rm ej}}{10 \, M_\odot} \right)^{0.21}
         \left( \frac{E_{\rm exp}}{10^{51} \, {\rm erg}} \right)^{-0.79}
         \left( \frac{R_0}{500 \, R_\odot} \right)^{2.16}.
\end{eqnarray}
Here the structure parameter $\rho_1/\rho_*$ in \citet{matzner99}
is taken to be unity, which is the case in our s17.0 model.
Our s17.0 model has an explosion energy 
of $E_{\rm exp} = 1.23 \times 10^{51} $ erg
at the end of the simulation.
Although we do not follow the hydrodynamics until shock breakout,
we assume that this is equivalent with the internal energy at the breakout.
Assuming that the outer envelope of the progenitor model is well-approximated
by a polytrope of index of $n=1.5$ and the opacity is dominated by
Thomson scattering ($\kappa = 0.34 \ {\rm cm^2\ g^{-1}}$), 
the expected luminosity of the shock breakout is 
$L_{\rm SBO} = 2.0 \times 10^{45}$ \ergs\
for $\Mej = 11.7 \Msun$ and $R_0 = 960 \Rsun$. Here, $\Mej$ is
the ejecta mass estimated as the progenitor mass right before
collapse minus the protoneutron star (PNS) mass. 
The duration of the emission is about 0.8 hour.

After the SBO emission, the
Type IIP supernova enters the characteristic plateau phase. 
The luminosity and duration time of the plateau phase
are given by \citet{popov93} as

\begin{eqnarray}\label{eq:Lplt}
L_{\rm plateau} & \sim & 1.6 \times 10^{42} \, {\rm ergs \, s}^{-1}  
                       \left( \frac{\kappa}{0.34\, {\rm cm^2 \, g}^{-1}} \right)^{-1/3} \nonumber \\
                      && \times \left( \frac{M_{\rm ej}}{10 \, M_\odot} \right)^{-1/2} 
                       \left( \frac{E_{\rm exp}}{10^{51} \, {\rm erg}} \right)^{5/6} \\
                      && \times   \left( \frac{R_0}{500 \, R_\odot} \right)^{2/3} 
                       \left( \frac{T}{5000 \, {\rm K}} \right)^{4/3}, \nonumber 
\end{eqnarray}
and
\begin{eqnarray}\label{eq:tplt}
t_{\rm plateau} & \sim & 99 \,{\rm d} 
                       \left( \frac{\kappa}{0.34 \, {\rm cm^2 \, g}^{-1}} \right)^{1/6} \nonumber \\
                      && \times \left( \frac{M_{\rm ej}}{10 \, M_\odot} \right)^{1/2} 
                       \left( \frac{E_{\rm exp}}{10^{51} \, {\rm erg}} \right)^{-1/6} \\
                       && \times \left( \frac{R_0}{500 \, R_\odot} \right)^{1/6} 
                       \left( \frac{T}{5000 \,{\rm K}} \right)^{-2/3}. \nonumber
\end{eqnarray}
Here, $T$ is the ionization temperature ($\sim$ 5000 K for hydrogen). After the termination of the plateau phase, the luminosity is approximated by the radioactive decay luminosity of $^{56}$Ni and $^{56}$Co with full trapping of gamma rays and positrons.


\bsp	
\label{lastpage}
\end{document}